%% file: ms.tex
\documentclass[iop]{emulateapj}

\usepackage{url}
\usepackage{psfig}
\usepackage{apjfonts}

\slugcomment{Submitted to \emph{The Astrophysical Journal
    Supplement Series}}

\shorttitle{Catalog of Rotation Measures and Redshifts}
\shortauthors{Hammond, Robishaw \& Gaensler}

\begin{document}

\bibliographystyle{apj_long_etal}

\title{A New Catalog of Faraday Rotation Measures and Redshifts \\ for
Extragalactic Radio Sources}

\author{Alison M. Hammond, Timothy Robishaw$^*$ \& B.~M.~Gaensler$^\dagger$}
\affil{\footnotesize{Sydney Institute for Astronomy, School of Physics,
    The University of Sydney, NSW 2006, Australia \\
$^*$ Covington Fellow, Dominion Radio Astrophysical Observatory, Herzberg
Institute of Astrophysics, National Research Council Canada, PO Box 248,
Penticton, BC, V2A 6J9, Canada \\
$^\dagger$bryan.gaensler@sydney.edu.au}}

\begin{abstract}

We present a catalog of Faraday rotation measures (RMs) and redshifts for
4003 extragalactic radio sources detected at 1.4~GHz,
derived by identifying optical counterparts and spectroscopic redshifts for
linearly polarized radio sources from the NRAO VLA Sky Survey. This catalog is more than an
order of magnitude larger than any previous sample of RM vs.\ redshift, and
covers the redshift range $0 < z < 5.3$;
the median redshift of the
catalog is $z=0.70$, and there are more than 1500 sources at redshifts
$z>1$.  For 3650 of these sources at Galactic latitudes $|b| \ge 20^\circ$,
we present a second catalog in which we have corrected for the foreground
Faraday rotation of the Milky Way, resulting in an estimate of the residual
rotation measure (RRM) that aims to isolate the contribution from
extragalactic magnetic fields.  We find no significant evolution of RRM with redshift,
but observe a strong anti-correlation between RRM and fractional
polarization, $p$, that we argue is the result of beam depolarization from
small-scale fluctuations in the foreground magnetic field or electron
density. We suggest that  the observed variance in RRM and the
anti-correlation of RRM with $p$ both require a population of magnetized
intervening objects that lie outside the Milky Way but in the foreground to
the emitting sources.

\end{abstract}


\section{Introduction}
\label{sec:introduction}

Faraday rotation is a powerful probe of magnetic fields along the
line of sight between a linearly polarized radio source and the observer. 
When combined with
redshift information, Faraday rotation measurements provide the
potential to constrain the evolution of cosmic magnetic fields
over time scales corresponding to the redshifts of the most distant
polarized sources detectable. However, previous studies of Faraday rotation
measure (RM) vs.\ redshift have been hampered by the small size of
the available samples (200--300 sources at most), often drawn from highly
inhomogeneous data sets.
Radio and optical surveys such as the NRAO Very Large
Array Sky Survey \citep[NVSS;][]{condon}, and the Sloan Digital Sky Survey
\citep[SDSS;][]{yaa+00} can enable
a comprehensive new analysis of RM
vs.\ redshift, thereby allowing much more sensitive studies of the evolution
of magnetic fields over cosmic time.

\subsection{Faraday Rotation}

Faraday rotation is a birefringence effect whereby the linear polarization
angle ($\theta$) of a radio wave is rotated when propagating through a
magnetized, ionized gas.  For a distant linearly polarized radio source
observed at a wavelength $\lambda$, the polarization angle is rotated by an
amount $\Delta \theta = {\rm RM}~\lambda^2$, where RM is defined as:
\begin{equation}
 {\rm RM}(z_{s}) = 0.81
 \int_{z_{s}}^{0} \! \frac{n_{e}(z)B_{||}(z)}{(1+z)^{2}} \frac{dl}{dz} dz
{\rm\ rad\ m^{-2}}.
 \label{eq:rm}
\end{equation}
In this Equation, $z_{s}$ is the redshift of the polarized source;
$n_{e}(z)$ is the free electron number density at some foreground redshift
$z$, in cm$^{-3}$; $B_{||}(z)$ is the line-of-sight component of the
magnetic field at redshift $z$, measured in $\mu$G; and $dl$ is a
line-element along the line of sight, measured in parsecs.  A positive RM
corresponds to a magnetic field oriented toward the observer.  In the
idealized case of a Faraday thin system, one can determine the RM by
measuring values for $\theta$ at multiple observing wavelengths, and then
performing a linear fit to $\theta$ vs.\ $\lambda^2$.

Equation (\ref{eq:rm}) implies that the observed RM is a superposition of
contributions from multiple magneto-ionic regions along the entire line of
sight. In every case, there is a contribution from the Earth's ionosphere
($\sim$$1$--$2 {\rm\ rad\ m^{-2}}$; \citealt{brentjens}), and for
extragalactic sources there is always a foreground contribution from the
Milky Way known as the Galactic Rotation Measure (GRM; $\sim$$10$--$1000
{\rm\ rad\ m^{-2}}$, varying with Galactic longitude and latitude;
\citealt{schnitzeler,oppermann}). Outside of the Galaxy, there are many
contributions to the observed RM, including the polarized sources
themselves, intervening galaxies, their halos, intracluster gas or
independent clouds of magnetized gas. The extragalactic component of RM can
be studied by calculating the residual rotation measure (RRM), such that RRM
= RM -- GRM.

\subsection{Previous Catalogs of RM and Redshift} 

Over the past several decades, there have been numerous studies in which
RRMs of polarized sources as a function of redshift have been used to probe
extragalactic magnetic fields over cosmic time \citep{bernet, fujimoto,
kronberg1976, kronberg1977, kronberg1982, kronberg2008, nelson, oren,
reinhardt, sofue, thomson, welter, you}. These studies have demonstrated
that the mean RRM of sources over the sky is zero out to redshifts $z > 4$.
However, the variance of the RRM distribution, the effect of
intervenors/absorbers and the precise nature of any RRM evolution with
redshift all remain contentious. The largest sample reported to date is the
(as-yet unpublished) data set analyzed by \citet{kronberg2008}, whose sample
consisted of 268 RRM measurements for sources out to $z \sim 3.7$. These
authors presented evidence for a significant increase in the variance of the RRM
distribution as a function of increasing redshift, from which they proposed
that galaxies hosted strong magnetic fields at relatively early times.

In the present paper, we use archival surveys and databases to derive a new
catalog of RM and redshift for more than 4000 extragalactic radio sources
out to redshifts $z \sim 5.3$, a sample that is markedly larger and extends
to much higher redshifts than all previous efforts.  We also present an
additional catalog in which we have corrected for Galactic Faraday rotation
for most sources, resulting in a compilation of RRM vs.\ $z$ suitable for
studies of the evolution of Faraday rotation and magnetic fields as a
function of cosmic time.  In \S \ref{sec:databases surveys} we describe the
databases and surveys used in our analysis.  In \S \ref{sec:catalog}, we
present our catalog of RM vs.\ $z$, while in \S \ref{sec:fgd} we present a
catalog of RRM vs.\ $z$. In \S \ref{sec:analysis} we discuss the overall
properties of the sample, and consider the relationships between RRM, fractional
polarization and redshift for these sources.

\section{Databases and Surveys}
\label{sec:databases surveys}

In the following subsections, we describe the radio and optical data sets
from which we obtained values of RM and $z$, respectively. The sky coverages
of the surveys used in this study are indicated in Figure \ref{fig:sky
coverage}; the final catalog contains only sources with declinations $\delta
\ge -40^\circ$, constrained by the coverage of the NVSS (from which all our
RMs have been derived).

\subsection{Obtaining Rotation Measures}
\label{subsec:taylor nvss}

The NVSS \citep{condon} was a 1.4 GHz radio survey that used the Very Large
Array (VLA) to image the entire sky for $\delta  \ge -40^{\circ}$ at an
angular resolution $R \approx 45''$.  The resulting source catalog contains
information on Stokes parameters $I$, $Q$ and $U$ for over 1.8~million
discrete objects.

The RMs that we use for our analysis come from the reprocessing of the
original NVSS data by \citet{taylor}, and are available
online.\footnote{\url{http://www.ucalgary.ca/ras/rmcatalogue}} \citet{taylor} used images of
Stokes $Q$ and $U$ in two frequency bands at 1364.9 MHz and 1435.1 MHz for
every NVSS source to determine linearly polarized intensity
$P\equiv\sqrt{Q^{2}+U^{2}}$ and fractional linear polarization $p\equiv
P/I$. If the signal-to-noise ratio for $P$ was greater than 8, they
derived the RM from the difference in polarization position angle across the
two bands. The resultant distribution of 37\,543 RMs is provided in Figure
\ref{fig:taylor_cat}. Because the RMs were derived only from two closely
spaced  frequency bands, the data are potentially subject to $n\pi$
ambiguities and non-$\lambda^2$ effects.  However, subsequent broadband
polarization studies for subsets of the \citet{taylor} catalog have produced
results that are in up to 96\% agreement with the RMs derived by
\citet{taylor} \citep{mao,vaneck,law}.

\subsection{Obtaining Redshifts}
\label{subsec:z_sources}

Our sources of redshift data are two online databases plus several recent
large optical redshift surveys.  We exclude from consideration any objects
flagged as stellar or Galactic. We also only use spectroscopic redshifts (as
opposed to photometric redshifts or Lyman dropouts) due to the greater
reliability and sufficient abundance of spectroscopic redshifts from the
surveys we consider. Details on these databases and surveys
are provided in \S\ref{subsubsec:ned} through \S\ref{subsubsec:2qz}.

To assign redshifts to the RM catalog of \citet{taylor}, we need to make
associations between sources detected at both optical and radio wavelengths.
There have been numerous approaches to this issue \citep{magliocchetti,
ivezic, mcmahon, best, sadler2007, kimball, lin, plotkin}.  Such studies use
angular proximity as the primary discriminant in making an association
between an optical source and a radio source, although the proximity
requirement varies widely from $2\arcsec$ to $3\arcmin$, depending on the
priority placed by different investigators on finding all genuine
associations or minimizing false matches. Some authors have considered the
possibility of multi-component sources with varying degrees of complexity,
from merely incorporating double-lobed radio sources to consideration of all
of double, triple, bent and core-jet scenarios \citep{magliocchetti, best}.
Matching procedures also differ based on whether associations are made with
a relatively low-resolution radio survey such as the NVSS ($R=45''$), a
high-resolution survey such as FIRST\footnote{The VLA Faint Images of the
Radio Sky at Twenty centimeters survey (see \S \ref{subsec:first} for more
details).} ($R=5''$; \citealt{becker, ivezic}), or a combination of both
\citep{best, plotkin}. Some procedures make use of properties such as source
size, flux or inclination between radio lobes \citep{magliocchetti, best},
while other methods calculate excess matches above those from random source
distributions \citep{kimball, plotkin}.

\subsubsection{NED}
\label{subsubsec:ned}

The NASA/IPAC Extragalactic Database
\citep[NED;][]{hms+91,hms+95}\footnote{\url{http://ned.ipac.caltech.edu/}~,
accessed on 2012~August~16.} stores data on $>$170~million objects outside
the Milky Way.  NED includes 4 million sources with redshift information,
including many associations between radio sources and optical redshifts.
These data come from surveys, published results and references to known
redshifts in the literature.  We discarded any redshift that was not
obtained spectroscopically, and also discard entries for which the ``$z$
Quality'' field was flagged.

\subsubsection{SIMBAD}
\label{subsubsec:simbad}

The Set of Identifications, Measurements and Bibliography for Astronomical
Data
\citep[SIMBAD;][]{woe+00}\footnote{\url{http://simbad.u-strasbg.fr/simbad/},~accessed
on 2012~August~17.}
is a database operated out of the Center de Donn\'{e}es astronomiques de
Strasbourg (CDS) in Strasbourg, France. It contains nearly five million
Galactic and extragalactic objects.  Like NED, SIMBAD includes redshift data
for radio objects derived from surveys and from the literature.  

\subsubsection{SDSS} \label{subsubsec:sdss} 

The SDSS\footnote{\url{http://skyserver.sdss3.org/dr8/en/}} \citep{yaa+00}
is an optical survey that has so far imaged more than one-third of the sky
(14\,500 square degrees) in five different wavebands at $R\approx
0.5\arcsec$, using a dedicated 2.5~m optical telescope at Apache Point, New
Mexico.  For the analysis in this paper we use data release 8 (DR8),
released on 2011 January 11 \citep{aihara}, which contains spectroscopic
data (including redshifts) for approximately 860\,000 galaxies and 116\,000
quasars over almost 9300~square degrees.  We used Structured Query Language
(SQL) to obtain data on spectroscopic redshifts and optical magnitudes for
galaxies and quasars from the online DR8 database.  We applied filters to
our search to select only objects classified as galaxies or quasars, and to
selectively remove any sources which had warnings associated with the
redshift measurements.

\subsubsection{The 6dFGS Survey}
\label{subsubsec:6df}

The Six-degree Field Galaxy Survey
\citep[6dFGS;][]{jones2004,jones}\footnote{\url{http://www-wfau.roe.ac.uk/6dFGS/}}
measured the spectroscopic redshifts of 110\,256 galaxies over 17\,000
square degrees, with a median redshift $z=0.053$.  The survey used the 6dF
fibre-fed multi-object spectrograph on the UK Schmidt telescope of the
Australian Astronomical Observatory (AAO). We applied filters to our search
to select only objects in the top two redshift quality bands, as determined
by the survey (quality 3 or 4).

\subsubsection{The 2dFGRS Survey}
\label{subsubsec:2df}

The Two-degree Field Galaxy Redshift Survey
\citep[2dFGRS;][]{colless2001,colless2003}\footnote{\url{http://www2.aao.gov.au/2dFGRS/}}
was carried out with the 2dF instrument on the Anglo-Australian Telescope (AAT). It covered 
approximately 1500 square degrees and obtained reliable redshifts 
for 221\,414 galaxies at a median redshift $z=0.11$. We applied filters
to select only objects in the superior redshift quality bands, as determined
by the survey (quality 3--5).

\subsubsection{The 2QZ/6QZ Survey}
\label{subsubsec:2qz}

The 2dF QSO Redshift survey (2QZ) and the 6dF QSO Redshift survey
(6QZ)\footnote{\url{http://www.2dfquasar.org/Spec\_Cat/catalogue.html} }
\citep{croom} were the quasar spectroscopic redshift survey counterparts,
respectively, to the 2dFGRS and 6dFGS described above. In combination they
are referred to as the 2QZ/6QZ survey. Combined, the 2QZ/6QZ survey contains
redshift data for 23\,660 quasars over 720 square degrees. We applied
filters to select only objects with high-quality redshifts, as determined by
the survey team (\texttt{Quality\_Flag\_1} = 11, 21 or 31).

\subsection{Morphological Information}
\label{subsec:first}

We additionally make use of the higher angular resolution of the FIRST
survey, with which we  match radio sources with optical data using
morphological information.  The FIRST survey \citep{becker} consists of VLA
images at 1.4~GHz covering 9900 deg$^2$, matched to the sky coverage of
SDSS.
FIRST detected  approximately one million sources, and its high angular resolution
($R=5''$ compared to $R=45''$ for the NVSS) enables the detection of compact
sources and provides detailed morphological information.

\section{The RM-Redshift Catalog}
\label{sec:catalog}

Previous catalogs of RM vs.\ $z$ such as those presented by \citet{welter}, \citet{oren} and
\citet{kronberg2008} have suffered from a high degree of
inhomogeneity. Typically, both RM and $z$ data have been compiled from a
range of small published (and occasionally
unpublished) samples, which were then collated into an RM-$z$ catalog. This
approach was undertaken to maximize the size of the data set, although 
the results were still only catalogs
of 100--200 sources. With the wealth of new data available,
we here attempt to strike a balance between creating a catalog with a
large number of sources, while also maintaining a high degree of
homogeneity in the data sets from which our information comes.
To associate optical redshift data with the \citet{taylor} RM catalog, we
first identify sources for which associations had previously been made in
NED or SIMBAD, and then use data from the optical redshift surveys
described in \S\ref{subsec:z_sources} to make new associations.

\subsection{NED and its Limitations}
\label{subsubsec:ned limits}

NED is currently the largest single database of extragalactic sources, and
therefore represents the natural starting point to search for redshifts. A
total of 2023 of the redshifts that we ultimately include in our RM-redshift
catalog were provided from information in NED (50.5\% of 4003
RM-redshift pairs). These represent cases for which a specific NVSS
source has an associated redshift within the NED database. 

Though many of the \citet{taylor} NVSS sources are listed within NED,
testing with the Sesame Name
Resolver\footnote{\url{http://cdsweb.u-strasbg.fr/cgi-bin/Sesame}} revealed
that some NVSS sources are missing. Correspondence with the NED team (M.
Schmitz, 2011, private communication) confirmed that only around 90\% of the
NVSS source catalog is in NED. The database is complete for right ascensions
$0^{\rm h} < {\rm RA} < 5^{\rm h} 40^{\rm m}$, but outside this range the
coverage is incomplete. Inquiries revealed that source inclusion is being
deferred until associations are made with preexisting entries in NED. This
task has been ongoing for three years, with no firm completion date planned.
Those sources not included have complications such as a single NVSS source
being resolved into several FIRST sources or the presence of more than one
object in the relevant field. Unfortunately, those sources not included are
precisely those that are especially relevant for this project, as they are
radio galaxies that have likely also been observed in the optical by
spectroscopic surveys. We therefore make these associations ourselves, as
described in \S\ref{subsec:association}.

\subsection{A Radio-Optical Association Algorithm}
\label{subsec:association} We add to the redshifts for NVSS sources provided
by NED and SIMBAD  by associating the radio sources in the \citet{taylor}
catalog with redshifts from the four optical surveys described in
\S\ref{subsec:z_sources}.  There are, however, a number of difficulties with
making associations between optical and radio data. One significant example
is galaxies which have complex radio morphologies, such as double-lobed
structures, that are not mimicked in the optical. To help make these
associations, we include data from the higher resolution FIRST survey
(though some NVSS sources either lack a FIRST detection or lie outside the
FIRST sky coverage). There is good complementarity between the high angular
resolution of FIRST and the superior sensitivity of NVSS to extended
emission \citep{best}, allowing detection of both compact and multi-component
radio sources.

Our association algorithm is primarily based on the work of \citet{best},
who tested and refined their criteria using Monte Carlo simulations
involving catalogs of random sky locations.  The association criteria
arising from these simulations, based on the entire NVSS catalog of 1.8
million
sources, were preferred to attempting to formulate new criteria using
simulations on the much smaller subset of 37\,543 source listed by
\citet{taylor}.  We also incorporate aspects of the schemes of
\citet{magliocchetti}, \citet{ivezic}, and \citet{mcmahon}; a summary of
the main elements of our algorithm is represented in Figure
\ref{fig:algorithm}.  Excluding associations taken from NED and SIMBAD, the
algorithm contributes 89\% of the RM-$z$ pairs in our RM-redshift catalog; the
remainder were flagged by the algorithm as having complex features and were
subsequently manually inspected. 

\subsubsection{Association Classes} 
\label{subsubsec:assc.classes}

Our algorithm associates sources and assigns them to one of five association
classes, A, B, C, D or E, as summarized in Table~\ref{tab:cat_classes}.
For sources in class~A, we have used FIRST data to make the association;
for sources in class~B, we rely on NVSS data only (the particular
limitations of class B are described in \S \ref{subsec:matching tests});
class C corresponds to more  complex sources that required
manual visual inspection; class~D corresponds to associations made
within NED, while class~E indicates associations made by SIMBAD.
Class~A is further subdivided into seven subclasses, A(i) through A(vii),
depending on the radio morphology of the source in the FIRST data.

The admission criteria for association classes A to C are further explained
below.  Examples of sources in each of classes A(i), A(ii), A(iii), A(iv),
A(v), A(vi), A(vii), B and C are provided in
Figures \ref{fig:example_associations1}, \ref{fig:example_associations2},
\ref{fig:example_associations3}, \ref{fig:example_associations4},
\ref{fig:example_associations5}, \ref{fig:example_associations6},
\ref{fig:example_associations7}, \ref{fig:example_associations8} and
\ref{fig:example_associations9}, respectively.

Following \citet{best}, we account for the possibility of multi-component
NVSS sources by making an initial search of the \citet{taylor} catalog in a
3$\arcmin$ radius around each optical source.  The distance of 3$\arcmin$ is
selected to ensure that genuine multi-component sources have at least two
matches, while still being much less than the typical separation of $8'-10'$
between NVSS sources. The chance of random matches within 3$\arcmin$ is
high, motivating more complex selection criteria as described below. As
shown in Figure \ref{fig:algorithm}, our scheme automates associations in
cases where there are fewer than three NVSS matches.

In cases for which there is one NVSS source within 3$\arcmin$ of a given
optical source, our algorithm checks for FIRST data within 30$\arcsec$ of
the optical position. If FIRST data are found and if the NVSS-optical offset
is less than 15$\arcsec$, the algorithm proceeds.  A single NVSS source more
than 15$\arcsec$ from the optical source is not considered an association.
The next steps taken by the algorithm depend on how many FIRST matches are
found within 30$\arcsec$.

A single FIRST match is accepted if the FIRST-optical offset is less than
$3\arcsec$ [class A(i); see examples in
Fig.~\ref{fig:example_associations1}] or if the FIRST-optical match is
within $10\arcsec$ and is also within 75\% of the major axis of the optical
source [class A(ii); Fig.~\ref{fig:example_associations2}]. The rate of
false detections for matches within 3$\arcsec$ is estimated to be less than
1\%, while the inclusion of matches within 10$\arcsec$, subject to the
additional criterion for class A(ii), helps maintain high completeness.

A double or triple match within FIRST is accepted if one of the two radio
sources is within $3\arcsec$ of the optical galaxy [classes A(iii) and A(v),
shown in Figs.~\ref{fig:example_associations3} and
\ref{fig:example_associations5}, respectively] or if the doubles criterion
is met for two of the matched
FIRST sources [classes A(iv) and A(vi), shown in
Figs.~\ref{fig:example_associations4} and
\ref{fig:example_associations6}, respectively). These require that the
quantity $O \times S \times F$ is less than five, where $O$ is the arcsecond
offset between the optical galaxy and the mean position of two FIRST
sources, $S$ is the ratio of source size for those two FIRST sources, and
$F$ is their flux ratio. This derives from the finding of \citet{mcmahon},
that double-lobed radio sources detected in FIRST have two lobes of similar
sizes and fluxes, and a mean position that is close to the optical galaxy.
The reliability of this procedure is estimated to be $>99\%$.

Our scheme also treats the case where two NVSS sources are found within
3$\arcmin$ of the same optical source. These sources are classified as
potential double-lobed radio systems if a set of offset criteria are met:
both radio sources must be within $90\arcsec$ of the optical galaxy, the
mean position of the two radio sources must match the optical position
within $15\arcsec$, and the nearer of the two radio components to the
optical source must be $>15\arcsec$ from the optical position (or there is
less than $20\arcsec$ between the two radio components). We then search for
FIRST data within 30$\arcsec$. \citet{best} did not consider sources in the
absence of FIRST data, but noted the likely degree of contamination in the
case of doubles if NVSS data alone are relied upon. For this work, the
\citet{taylor} catalog contains sufficiently few candidate doubles that
those without clear FIRST data (i.e., a FIRST source within 3$\arcsec$ of
the optical source position) are visually inspected (see below). Potential
doubles with clear FIRST data are associated as doubles and placed in class
A(vii). Note that sources in class A(vii) have two NVSS sources (the two
lobes of a radio galaxy) matched to the same optical source.  Within the
RM-redshift catalog, the two components of each pair are arbitrarily
designated class A(vii)a or A(vii)b in reflection of this.
Figure~\ref{fig:example_associations7} shows two examples of class A(vii)
associations.

In some cases for which there is only one NVSS match, there are no
corresponding FIRST data because the source lies outside the sky coverage of
the FIRST survey (see Figure \ref{fig:sky coverage}), or because the source
has faded between the NVSS and FIRST observations, or because a radio source
seen as substantially extended in NVSS is resolved out in the higher angular
resolution FIRST data.  In these cases, a match is accepted if the
NVSS-optical offset is less than 10$\arcsec$, and the source is then placed
in class B. Figure \ref{fig:example_associations8} shows two examples of
class B sources. The choice of a 10$\arcsec$ matching radius is conservative
in light of other work, and avoids significant contamination from background
sources \citep{best, plotkin, sadler2007}. It is also smaller than the
NVSS-optical offset limit of 15$\arcsec$ that we use for class A (i.e., when
a FIRST counterpart is detected).

For sources for which there are three or more NVSS matches, four or more
FIRST matches, or a candidate double source with no clear FIRST data, the
algorithm flags the optical source as requiring manual visual inspection. In
these complex situations, images of the relevant parts of the sky are
downloaded from the NVSS and FIRST postage stamp servers, the SDSS image
server and the SuperCOSMOS image extraction service (for 2dFGRS, 6dFGS and
2QZ/6QZ). In cases for which we confirm visually that the optical and radio
sources are associated, then the value for redshift from the optical
survey is assigned to the \citet{taylor} source, and these sources are
assigned class C. Two examples of class C sources are shown in
Figure~\ref{fig:example_associations9}. While visual inspection introduces
some level of subjectivity to the resulting associations, we note that only
195 associations in our final RM-redshift catalog fall into this class, just
4.9\% of the total.

\subsubsection{Noteworthy Features of the Association Algorithm} 
\label{subsubsec:int.feat}

In 25 cases (23 of which were drawn from 6dFGS), the algorithm matches two
different sources from the same optical survey with  a single radio source
from \citet{taylor}. In all such cases, we visually inspected
the optical and radio images, using morphology and proximity to
determine which optical source is associated with
the radio source.

We note that while our association scheme attempts to use NVSS and FIRST
together to achieve the best possible compromise between completeness and
reliability, our approach does exhibit a slight bias against extended
sources. This is because nearly all matches involving a single-component
FIRST source are detected by our algorithm, while not all of those involving
multiple NVSS components are identified  \citep{best}. There is no {\em a
priori}\ reason that this slight bias should impact the results of our RM
analysis.

\subsection{Choosing a Final Redshift for the RM-Redshift Catalog}
\label{subsec:bestz}

Many of the polarized sources in the \citet{taylor} catalog have an optical
counterpart and resulting redshift in more than one of the optical surveys
or databases that we have considered.  We list every such redshift in our
RM-redshift catalog, but additionally identify a ``selected redshift'' for each
\citet{taylor} source, representing that redshift which we consider to be
most reliable.  For example, 59 NVSS sources have redshifts from two
separate optical surveys, while one source was associated with three optical
surveys. In 2296 cases, associations of \citet{taylor} sources are made
with both databases and surveys (for example, a \citealt{taylor} source that
it is associated with an SDSS optical counterpart by our algorithm will
sometimes have the same NVSS/SDSS match listed in NED), or an optical
counterpart is listed both in NED and in SIMBAD.  In many cases, the
decision on a selected redshift is trivial, either because all redshift
entries are the same (730 cases), or because all the associated redshifts
agree closely with one another (1439 cases with $0 < \Delta z \le 0.01$). From
our total of 4003 sources, this leaves 127 cases for which multiple distinct
redshifts are found, so that the selection of a final redshift could have a
functional impact upon the catalog. The differing redshifts are due to
disagreements between the constitutive surveys/databases. Some of these
disagreements are attributable to identifiable errors such as measuring the
redshift of an absorber along the line of sight to an emitter and wrongly
assigning that redshift to the emitter.

In those cases where a substantive redshift selection needs to be made, the
final selected redshift is chosen from among detections in multiple optical
surveys/online catalogs in the descending priority listed in
Table~\ref{tab:cat}. This prioritization gives preference to surveys over databases,
and prefers surveys with high angular resolution and wide-area sky coverage.
NED is preferred to SIMBAD because of the greater number of reliable
redshifts in NED.

\subsection{Contents of the RM-Redshift Catalog}
\label{sec:contents}

As described in \S\ref{subsec:z_sources} and \S\ref{sec:catalog}, we have
obtained redshifts for the polarized radio sources of \citet{taylor} from a
range of optical databases and surveys, with each optical/radio match given
a corresponding association class as defined in Table~\ref{tab:cat_classes}.
The resulting RM-redshift catalog contains 4003 sources with both RMs and
redshifts (with a ``selected'' redshift listed in cases where multiple
values for $z$ have been found).  This catalog contains an
order-of-magnitude more entries than any previous compilation of RM vs.\
redshift, and extending out to much higher redshifts ($z=5.27$). A sample of
rows and columns from the RM-redshift catalog is shown in Table
\ref{tab:rm_z_catalog}; the full RM-redshift catalog (available in its
entirety as an online machine-readable table) contains the coordinates,
fractional polarization, RM and all other data for each source from
\citet{taylor}, along with positions, redshifts, photometry, object types,
journal references and association class derived for each optical match,
and the selected redshift and other relevant data associated with the
best-matching optical counterpart.

Table \ref{tab:sdss} shows the contents of the RM-redshift catalog broken
down by survey and association class. The most common association class is
A(i), representing a simple unresolved radio source that aligns closely with
the corresponding optical counterpart.  This is followed by class B,
representing a similar situation to class A(i) but in cases either where no
FIRST source is detected or no FIRST data are available.
The 2dFGRS, 6dFGS and 2QZ/6QZ surveys provide most of their
associations in class B, 
as their sky coverage has little overlap overlap with FIRST. In contrast,
almost all SDSS associations fall in class A, because the sky-coverages of
FIRST and SDSS were intentionally matched.  The total number of associations
arising from each survey varies as expected according to the sky coverage
and number of objects in each survey. For example, most survey associations
are made with the SDSS, covering 9300 square degrees and containing almost
1 million extragalactic redshifts, while the fewest associations are made
with 2QZ/6QZ, containing only 23\,000 quasar redshifts and focusing on two
small regions of the sky.
The row labeled ``Total'' in Table~\ref{tab:sdss} indicates the total
number of matches to \citet{taylor} for each survey or database, 
while the column labeled ``All'' indicates the total number of
times each association class is assigned to a radio/optical match.
The rows and columns labeled ``Selected'' show the contribution
from each survey/database and from each association class, respectively,
to the final selected redshifts using the priority order listed in Table~\ref{tab:cat}.

\subsection{Completeness, Reliability and Testing of Matching Processes}
\label{subsec:matching tests}

It is desirable to estimate both the completeness and the reliability (also
known as efficiency; cf.\ \citealt{kimball}) of our RM-redshift catalog. Completeness
refers to the fraction of real matches found, while reliability refers to
the fraction of matches that are real. Both these quantities are difficult
to estimate for our RM-redshift catalog because they are reliant upon a number of
different factors: the completeness and proportion of spurious sources
within each constituent survey, the number of true matches expected, and
the level of background contamination in different regions of the sky.

Previous studies involving matches of NVSS with FIRST have found high
completeness and reliability \citep{best, kimball}.  \citet{kimball}
considered a range of matching radii;
for the radius of
$30\arcsec$ that we adopt, they found a completeness of 99.7\% and a reliability of
96\%. Matching between FIRST and SDSS yields a completeness above 94\% and
reliability above 90\%. This rises further when NVSS data are included, with
\citet{best} suggesting 95\% as a conservative completeness estimate.

Figure \ref{fig:offsets} shows the difference in NVSS and optical sky
positions for matched radio-optical sources. The distributions of offsets in
both Right Ascension and Declination are sharply peaked around zero, with a
small spread. More than 90\% of matched NVSS sources lie within $3\arcsec$
of the associated optical source; when double-lobed radio sources are
excluded, this fraction rises to 94\%.

One particular issue is the 375 sources in the RM-redshift catalog in
association class~B, i.e., sources that were associated by our algorithm
using optical and NVSS data only, without a FIRST detection. Class~B
associations are likely to be less reliable than those that also draw on
FIRST data. Specifically, \citet{best} have used Monte Carlo simulations to
estimate that this association class may have a false detection rate of up
to 6\%. However, class~B represents less than 10\% of the entries in the
RM-redshift catalog. Removing these sources would cause the completeness of
the catalog to suffer, and also would further bias the sample against
extended sources (due to the removal of extended sources that lie within the
FIRST sky coverage but lack a FIRST detection). For the RM-redshift catalog
as a whole, the simulations of \citet{best} suggest that schemes that use
both NVSS and FIRST have an overall reliability $>98\%$.

We tested our algorithm using the prior radio-optical association work of
\citet{oren} and \citet{kimball}. \citet{oren} provided a small test sample
of 20 sources that we used to test the basic functionality of our scheme. A
more robust test involves the larger catalog of \citet{kimball}, who
associated the SDSS with a range of radio surveys, including the
NVSS.\footnote{The \citet{kimball} catalog is available online at
\url{http://www.astro.washington.edu/users/akimball/radiocat/}} Of the
37\,354 \citet{taylor} sources, 1295 appear with redshift data in the work
of \citet{kimball}.

The \citet{kimball} catalog used SDSS DR6 \citep{adelman}, so we also performed tests using
that data set, downloaded from the SDSS archive. Using the matching radii
selected by \citet{kimball}, which were different from our own due to the differing
  aims of that study, we were able to make 1260 of the 1295
associations made by \citet{kimball}. The small differences are
attributable to different correlation programs (the \citealt{kimball}
scheme, for instance, takes the FIRST data, not the optical data, for the
position of the source), rounding differences arising from computation
using different programs, and oddities within the \citet{kimball} catalog. 

We also tested the robustness of our approach by randomly selecting 100
sources that had been associated via our algorithm, and visually inspecting
them. In all cases, the visual inspection confirmed the association made by
the code. This indicates that our algorithm associates optical and radio
detections with a reliability $>99\%$.

\section{The RRM-Redshift Catalog}
\label{sec:fgd}

\subsection{Subtraction of the Galactic RM Contribution}
\label{subsec:GRM}

In considering the evolution of magnetic fields over cosmic time, it is
insufficient to simply plot RM against redshift because the foreground
Faraday rotation from our own Milky Way must first be accounted for.  This
is especially important because the Galactic contribution is not uniform,
but varies across the sky.

Due to the increased spatial fluctuations in the GRM at low Galactic
latitudes, the first step in calculating RRM is to exclude all sources with
$|b|<20^{\circ}$ (drawing upon the study of these fluctuations by
\citealt{schnitzeler},
and taking a threshold for $|b|$ comparable to that
chosen for studies by \citealt{welter} and \citealt{oren}). We thus
only calculate an RRM for
3650 of the 4003 sources presented in \S\ref{sec:contents}, the breakdown
of which for each optical survey/database and for each association class is listed in
the final row and final column of Table~\ref{tab:sdss}, respectively.

A range of methods have been employed in the literature for calculating
and subtracting the GRM. In this paper, we use the new map of the
foreground Faraday sky computed by \citet{oppermann}, as shown in Figure 
\ref{fig:foreground_niels}. This map 
enables the calculation of an RRM by subtracting
the GRM at that sky position from the RM value reported by
\citet{taylor}. The \citet{taylor} RM data were used
in the construction of the \citet{oppermann} map, but the reconstruction
algorithm used in that work filters out the extragalactic contributions,
ensuring that the resulting map is still appropriate to use in GRM
calculations for \citet{taylor} sources. 

Figure \ref{fig:redshift_methods} compares RM and RRM for our data. The
efficacy of this GRM subtraction is indicated by the considerable narrowing
of the RRM distribution (mean $+0.3\pm0.4$~rad~m$^{-2}$,
standard deviation  23.2 rad~m$^{-2}$) compared to
the RM distribution (mean $+2.3 \pm 0.6$ rad~m$^{-2}$,
standard deviation 36.1~rad~m$^{-2}$). We also
considered a number of previously published RRM schemes
\citep{oren,johnston-hollitt,sofue,short}, but the GRM map of
\citet{oppermann} resulted by far in the smallest scatter in RRM of all
approaches considered.

\subsection{Contents of the RRM-Redshift Catalog}
\label{subsec:availability}

After applying the GRM correction discussed in \S\ref{subsec:GRM}, we derive
a catalog of 3650 radio sources with both RRMs and redshift.  An extract of
some columns from this RRM-redshift catalog can be seen in Table
\ref{tab:rrm_z_catalog}. The full RRM-redshift catalog (available online as
a machine-readable table),
contains the coordinates, flux, fractional polarization, RM, GRM, RRM,
redshift, plus appropriate ancillary data for the radio source from \citet{taylor} and for
the best-matching optical counterpart.

This RRM-redshift catalog is more than an order of magnitude larger than any
such catalog previously published. This is highlighted by Figure
\ref{fig:catalog_comparison}, in which we compare our catalog data
against the samples of RRM vs.\ $z$ described by \citet{kronberg1982},
\citet{welter} and \citet{oren}.

\section{Analysis of the RRM-Redshift Catalog}
\label{sec:analysis}

\subsection{Summary of Source Characteristics} 

The RRM-redshift catalog of 3650 sources covers a redshift range from 0
to 5.27, with a median redshift $z=0.70$ and including almost 1400 sources
with redshifts $z>1$. The RRM-redshift catalog covers an RM range from
$-465.4$ to $+270.4$ rad$\ $m$^{-2}$, and an RRM range from $-476.5$ to
$+206.1$
rad$\ $m$^{-2}$. The RRM distribution, however, is tightly clustered around zero:
88\% of sources have $|{\rm RRM}| < 25$ rad$\ $m$^{-2}$, 96\% have $|{\rm
RRM}|<50$ rad$\ $m$^{-2}$, and 99.3\% have $|{\rm RRM}|<100$ rad$\
$m$^{-2}$.  The 1.4 GHz Stokes~$I$ fluxes of the sample ranges from 11~mJy to 55~Jy,
with a median of 300~mJy. Polarized fluxes range from 2.6~mJy to 1.3~Jy,
with a median of 8~mJy.

The RRM-redshift catalog contains a diverse range of sources, drawn from
several surveys and databases, and with a range of multi-wavelength
properties and environments. A full analysis of this data set is beyond the
scope of this paper, but here we present a brief analysis of its overall
properties. To address potential issues of inhomogeneity in the catalog, we
consider the full data set plus three relatively large and well-defined
subsets: the 1376 sources for which the selected redshift was drawn from
SDSS (referred to as ``SDSS sources'' in subsequent discussion), the 516
sources with SDSS redshifts for which the optical spectrum was classified by
SDSS as a galaxy (``SDSS galaxies''), and the 860 sources with SDSS
redshifts for which the optical spectrum was classified by SDSS as a quasar
(``SDSS quasars'').

For each of these four data sets, we consider three relationships: RRM as a
function of $z$; fractional linear polarization, $p$, as a function of $z$;
and RRM as a function of $p$.  We note that any relation between RRM and $p$
does not explicitly require knowledge of the redshift of the sources as has
been the focus of this paper. However, because we have optical
identifications, consideration can be given to how the RRM vs.\ $p$ relation
changes for different subsets of the overall data such as SDSS galaxies vs.\
SDSS quasars.

%In the following subsections, we consider each of the three relationships
%between RRM, $p$ and $z$ in turn.

\subsection{Residual Rotation Measure vs.\ Redshift}
\label{sec_rrm_z}

Figure~\ref{fig_rrm_z} shows the distribution of RRM as a function of
redshift for (a) the entire RRM-redshift catalog (this is the same data as
shown in the bottom panels of Figs.~\ref{fig:redshift_methods} and
\ref{fig:catalog_comparison}), (b) all SDSS sources, (c) SDSS galaxies, and
(d) SDSS quasars.  As expected, there is a marked difference in the redshift
distribution of SDSS galaxies (median redshift $z=0.17$) compared to SDSS quasars
(median redshift $z=1.22$). 

As indicated by the red lines overlaid on each panel of
Figure~\ref{fig_rrm_z}, there is no apparent trend in the mean or standard
deviation of RRM as a function of $z$ in any of the four samples
considered.\footnote{The two places in Figure~\ref{fig_rrm_z}(a) where the
standard deviation is substantially larger than $\approx$20---25~rad~m$^{-2}$
are produced entirely from two sources with anomalously large RRMs (a source
at $z = 1.44$ with RRM~$=-444$~rad~m$^{-2}$ and a source at $z = 2.13$ with
RRM~$=-476$~rad~m$^{-2}$, respectively). The latter source is an SDSS
quasar, and so also has an effect on the standard deviations in panels (b)
and (d).}  This is in strong contrast to the results of 
\citet{kronberg2008}, who presented a significant increase in the standard 
deviation of RRM to higher redshifts. 

\citet{kronberg2008} further characterized an evolution of RRM with $z$ by
splitting their data into two groups, corresponding to sources above and
below a threshold redshift $z_b$. By applying a Kolmogorov-Smirnov (KS) test
to the normalized cumulative distributions of $|{\rm RRM}|$ at redshifts
above and below $z_b$, \citet{kronberg2008} found that the RRMs of low- and
high-redshift sources differed at 99\% significance for $z_b \sim 1.8$.  We
can repeat this experiment with far larger sample sizes: in our RRM-redshift
catalog, there are 3140 and 510 sources at $z_b < 1.8$ and $z_b \ge 1.8$,
respectively. Applying a KS test to the distributions of $|{\rm RRM}|$ on
either side of this threshold, the values  of $|{\rm RRM}|$ for the two
samples differ at 53\% significance, which is consistent with the two groups
of data being drawn from the same underlying distribution.

As a more sensitive test of any evolution of RRM with $z$, we use the Spearman rank
test to look for evolution of $|{\rm RRM}|$ as a function of $z$ for all
four data sets shown in Figure~\ref{fig_rrm_z}.  
We find no correlation for panel (c), and find 
a weak
correlation between  $|{\rm RRM}|$ and $z$, at 2.5$\sigma$, 2.2$\sigma$
and 3.8$\sigma$ significance for panels (a), (b) and (d),
respectively.  We do not consider any of these trends significant,
especially given that here and in further subsections below
we are examining a diverse range of different possible correlations.

We thus conclude that there is no significant dependence of RRM or its
variance as a function of $z$ in our RRM-redshift catalog, in contrast to
the strong effect of this kind reported previously using much smaller data
sets \citep{welter,you,kronberg2008}.  Recently, \citet{bml12} have considered a
much smaller sample of 371 RMs from \citet{taylor}, and reported a similar
inability to reproduce the RRM vs. $z$ behaviour seen by
\citet{kronberg2008}.  Given this lack of redshift dependence, we now
briefly consider possible origins of the observed RRMs.

We first note two  terms that must contribute to the standard deviation of
23.2~rad~m$^{-2}$ seen for the RRMs in Figure~\ref{fig_rrm_z}(a): the
measurement errors of the individual RMs in the \citet{taylor} catalog, and
the error associated with the GRM calculation of \citet{oppermann}.  The
standard deviation in RM due to measurement errors is 11~rad~m$^{-2}$
\citep{schnitzeler,sts11}, while the mean error in GRM for the sources in
our RRM-redshift catalog is 6~rad~m$^{-2}$. Subtracting these in quadrature from
the observed variance leaves a standard deviation of 20~rad~m$^{-2}$ that
must come from one or more astrophysical phenomena.

An obvious explanation for the lack of redshift evolution is that the RRMs
are solely a residual contribution from the Milky Way, resulting from imperfect
foreground subtraction, or from RM fluctuations on smaller angular scales
than are being sampled by the GRM map of \citet{oppermann}. We can rule out the
former option, because while the RMs of our sources show a strong dependence
on Galactic longitude and latitude (as seen in Fig.~\ref{fig:taylor_cat} for
the larger NVSS RM sample from which our catalog is derived), the
corresponding RRMs show no pattern or trend with Galactic coordinates.  To
consider the latter option, we note that the typical spacing between the
sources that \citet{oppermann} used to calculate their foreground map is
$\sim1^\circ$, meaning that any Galactic contribution to our observed RRMs
must represent GRM fluctuations on angular scales $\la 1^\circ$.
\cite{sts11} calculate the fluctuations in GRM on these scales and show
that, at latitudes $|b| \ge 20^\circ$ as we are considering here, the
standard deviation in RM on a scale of $1^\circ$ is $12-17$~rad~m$^{-2}$.
Subtracting this Galactic term in quadrature from our observed variance
implies that there is a 10--15~rad~m$^{-2}$ contribution to the standard
deviation in RRM that cannot be due to small-scale fluctuations in GRM, and
hence must be extragalactic. This can be compared with the work of
\citet{schnitzeler}, who performed a statistical decomposition of the RMs in
the \citet{taylor} catalog into different components as a function of
Galactic latitude, and concluded that extragalactic Faraday rotation
contributed a standard deviation of $\approx6$~rad~m$^{-2}$ to the
RMs of \citet{taylor}.  An agreement
within a factor of $\sim2$ between our estimate and that of \citet{schnitzeler}
seems reasonable given the very different approaches taken between the two
studies.

We now consider the possibility that most of the extragalactic contribution
to RRM arises in the polarized sources themselves, e.g., in an envelope of
ionized gas in the host galaxy or in its immediate environment.  If the RRM
originates at the same redshift as the emitter, then we expect a
$(1+z)^{-2}$ dilution factor as per Equation~(\ref{eq:rm}). The corresponding
contribution to the variance in RRM should then decrease with redshift in
Figure~\ref{fig_rrm_z} \citep[cf.\ the dashed line in Fig.~7
of][]{kronberg2008}.  Specifically, if we assume that the extragalactic
contribution to the standard deviation in RRM of 10--15~rad~m$^{-2}$
corresponds to sources at our median redshift $z = 0.7$, then an
identical population of sources at a
redshift $z=2$ should only contribute a standard deviation in RRM of
1--2~rad~m$^{-2}$. Considered in quadrature with the other terms
contributing to the scatter in RRM as discussed above, this dilution should
result in an overall decrease in the standard deviation of RRM by
2--4~rad~m$^{-2}$ between $z = 0.7$ and $z = 0.2$. Although this is a small
effect, we can rule out its presence in our data --- the Spearman rank test
discussed above shows that if anything, the variance in RRM slightly
increases, not decreases, with redshift.  The lack of evolution of RRM with
$z$ can then only be explained if the standard deviation in RRM evolves with
$(1+z)^2$ in the emitter's reference frame, to cancel out the $(1+z)^{-2}$
effect as the signal propagates to Earth. This would be a fortuitous
coincidence and would also represent very strong evolution of the emitted
RRM with redshift, both of which make this possibility unlikely.

The alternative is that the extragalactic component of RRM towards each
source is introduced in one or more intervening systems along the line of
sight between the polarized source and the observer, a possibility
considered in detail by many previous authors
\citep[e.g.,][]{bernet,kronberg2008,oren,welter}.  If the intervening
systems are all at comparable redshifts, then we will not see a $(1+z)^{-2}$
dilution term, and indeed might expect a slight increase in variance of RRM
with increasing $z$, since more distant sources are more likely to have an
intervenor along the line of sight \citep[see solid and dotted lines in
Fig.~7 of][]{kronberg2008}. \citet{kronberg2008} advocate a population of
intervenors with a standard deviation of RRM of $\sim$60--115~rad~m$^{-2}$
in the observer's frame, but our analysis argues that a contribution
10--15~~rad~m$^{-2}$ is more likely. This is broadly consistent with the
halo of a Milky-Way-like galaxy \citep[$n_e \approx 3\times10^{-4}$~cm$^{-3}$, $B
\approx 1$~$\mu$G, $l \approx 50$~kpc;][]{gmcm08,mao,sr12}, but a proper
interpretation requires further, detailed consideration of this and other
possible intervening source populations.

\subsection{Polarized Fraction vs.\ Redshift}
\label{sec_p_z}

Figure~\ref{fig_p_z} shows how $p$ varies with $z$ for the entire
RRM-redshift catalog, for all SDSS sources, for SDSS galaxies, and for SDSS
quasars. Panels (a) and (b) show a clear trend, in that sources with $z \la
0.5$ can have a range of polarized fractions extending beyond $p = 30\%$,
while sources at $z \ga 0.5$ are confined solely to lower polarized
fractions, $p \sim 5\%$.

Panels (c) and (d) provide a simple explanation for this apparent
bimodality. Within the SDSS sub-sample, the low-redshift sources that can be
both weakly or strongly polarized are all classified as galaxies based on
their optical spectra. In contrast, the SDSS sources that extend to high
redshifts and that are all relatively weakly polarized all have optical
spectra indicating that they are quasars. While this distinction needs
further study, a likely explanation is that the NVSS counterparts to the
SDSS galaxies are radio lobes from active galaxies, which are expected to be Faraday thin
and thus show high degrees of polarization. In contrast, the NVSS
counterparts to SDSS quasars are radio-loud cores, for which optical depth
effects and strong magnetic fields are expected to result in reduced polarization levels.

We can confirm this hypothesis by considering the distribution of
association classes for the SDSS galaxies compared to the SDSS galaxies. As
summarized in Table~\ref{tab:cat_classes}, the most robust classifications
are class A, in which an NVSS source has a clear match with an optical
counterpart and also shows a relatively simple morphology in FIRST. Within
class A, sub-classes A(i) and A(ii) are characteristic of core-dominated
radio morphologies, while classes A(iii) through A(vii) represent radio
morphologies for which a significant fraction of the emission is from radio
lobes. If the radio polarization from SDSS galaxies is mainly from radio
lobes and that from SDSS quasars is mainly from radio cores, then we expect
the SDSS galaxies to be dominated by classes A(iii) to A(vii), while the
SDSS quasars should be dominated by classes A(i) and A(ii).  Indeed the data
support this expectation: considering only SDSS sources in class~A, 35\% of
SDSS galaxies are in classes A(i) and A(ii), while 65\% are in classes
A(iii) through A(vii). In contrast, 54\% of SDSS quasars are in classes A(i)
and A(ii), while 46\% are in classes A(iii) to A(vii). Furthermore, we note
that class~C corresponds to complex, extended morphologies that cannot be
classified automatically. For SDSS galaxies, 20\% of all sources in the
RRM-redshift catalog are in class~C, while for SDSS quasars, the fraction of
class~C sources drops to just 8\%.  All this additional morphological
information clearly demonstrates that SDSS galaxies are dominated by lobes,
while SDSS quasars are dominated by cores, consistent with the differing
distributions of fractional polarization for these two populations seen in
Figure~\ref{fig_p_z}.

Noting the differing polarization properties of galaxies and quasars, we can
then consider whether there is any evolution of $p$ with $z$ in
Figures~\ref{fig_p_z}(c) and (d). Applying the Spearman rank test, we find
no significant correlation of $p$ with $z$ for either SDSS galaxies or SDSS
quasars, as apparent by eye via the red lines marking the mean and standard
deviation of $p$ as a function of $z$. A similar absence of evolution of $p$
with $z$ at 1.4~GHz was recently found by \citet{bgt+11} from a smaller sample of 69
sources in the redshift range $0.04 < z < 3.2$.

Other studies of polarization as a function of redshift have usually
considered parameters such as the ``depolarization measure'' (DP; the ratio
of $p$ at a high frequency to that at a low frequency) or $\lambda_{1/2}$
(the wavelength at which $p$ falls to half its peak value)
\citep[e.g.,][]{kcg72,goodlet}.  Many of these studies have found that DP
increases with $z$ or that $\lambda_{1/2}$ decreases with $z$, in both cases
indicating increased depolarization at higher redshifts
\citep{chk+74,goodlet,gk05,it82,kcg72}. While we do not see such an effect here, it
is important to note that such studies convert observed values of DP or
$\lambda_{1/2}$ to those seen in the reference frame of the emitter.  We
cannot make such a correction for our sample because we do not have
multi-wavelength information that would tell us how $p$ depends on
wavelength for each source. With the exception of the study of
\citet{bgt+11}, our result is thus not directly comparable with previous work
on the depolarization of radio sources as a function of redshift, without the
inclusion of several additional assumptions.

\subsection{Residual Rotation Measure vs.\ Polarized Fraction}
\label{sec_rrm_p}

In Figure~\ref{fig_rrm_p} we plot the dependence of RRM on $p$ for
the entire RRM-redshift catalog, for all SDSS sources, for SDSS galaxies, and for SDSS quasars.
In this case we see a striking pattern in all four panels: the
standard deviation in RRM is large for low fractional polarizations, and is small for
high fractional polarizations. 	

We first note that this is not an artifact of sensitivity and
signal-to-noise. The uncertainty in an individual RM measurement (and hence
the variance in a large sample of similar RM measurements) is inversely
proportional to the signal-to-noise ratio \citep{brentjens}. Thus we expect a source
with a low polarized flux to have a large error in RM,  and vice versa.
However, we are here plotting the fractional polarization rather than the
polarized flux, and there is no correlation between these two parameters in
our catalog. (If anything, any bias is in the reverse direction, since the
faintest sources can only be detected in polarization if their fractional
polarization is high.) The behavior seen in Figure~\ref{fig_rrm_p} is thus
astrophysical, rather than instrumental.

The most likely interpretation for this behavior is a
depolarization mechanism, since stronger Faraday effects in or in front of
a polarized source can induce reduced polarization through a variety of
mechanisms \citep[e.g.,][]{burn,gardner,tri91,sbs+98}.
Bandwidth depolarization cannot be a contributor in Figure~\ref{fig_rrm_p},
since this effect is only significant for RMs with magnitudes larger than
$\sim100$~rad~m$^{-2}$ \citep[see Fig.~1 of][]{taylor}.  Just 2\%
of the sources in our RRM-redshift catalog have $|{\rm RM}| >
100$~rad~m$^{-2}$, so this cannot be a significant contributor to the
observed depolarization. Depth depolarization is also unlikely
to be a factor, since this only occurs when the emitting medium
is mixed with the medium producing the Faraday rotation. We discussed in
\S\ref{sec_rrm_z} above how the lack of evolution of RRM with $z$  argues
against an intrinsic origin for the RRMs, meaning that the RRM signal occurs
wholly
in the foreground to the polarized sources and so cannot produce depth
depolarization.

The remaining possibility is beam depolarization, whereby high values of RRM
imply large fluctuations in RRM on scales smaller than the angular extent of
the radio source (even in cases where the source is unresolved by the
telescope beam). Such fluctuations will cause differing polarization angles
within the beam to cancel, resulting in a reduced polarized fraction. 

\citet{hbgm08} have shown that the Galactic foreground can indeed
simultaneously produce larger RMs and enhanced beam depolarization against
unresolved polarized background sources.  Beam depolarization due to the
Galactic foreground produces an anti-correlation between GRM and $p$
\citep{hbgm08}, and so thus can possibly produce a dependence between RRM
and $p$ given that one component of RRM is likely to be a residual Galactic
contribution produced by small-scale fluctuations in GRM (see
\S\ref{sec_rrm_z}).  While the depolarizing effects of the Galaxy are likely
to be present in our data, they are insufficient on their own to explain the
trend that we see between $|{\rm RRM}|$ and polarized fraction. For example,
Figure~6 of \citet{hbgm08} shows that for similar angular resolution and
observing frequency to those of the NVSS data being considered here,
fluctuations in Galactic RM at the level of $\sim$200~rad~m$^{-2}$ are
needed to depolarize background sources by a factor of $\sim2$. In contrast,
Figure~\ref{fig_rrm_p} suggests that fluctuations in RRM only of magnitude
$\sim50$~rad~m$^{-2}$ can depolarize background sources by a factor of
$\sim$3--5. The depolarization that we observe is thus too strong an effect
to be explained only by small-scale Galactic RM fluctuations.  The beam
depolarization also does not seem to be originating in the emitting sources
themselves, since we have argued in \S\ref{sec_rrm_z} above that the RRMs do
not have a significant intrinsic component.  A significant contribution to
the observed depolarization must therefore be due to small-scale
fluctuations in RM somewhere between the source and the Milky Way, in
intervening magneto-ionic material along the line of sight. This is the same
origin independently proposed in \S\ref{sec_rrm_z} for the extragalactic
component of the RRMs themselves. \cite{bml12} have used a different
approach, focusing on the frequency dependence of depolarization, but have
separately come to a similar conclusion that the extragalactic intervenors
contributing to the observed RRMs must be highly turbulent.

\section{Conclusions}
\label{sec:summary}

We have compared the extragalactic rotation measure catalog of
\citet{taylor} with redshift data from a range of optical surveys and
databases to produce a new sample of 4003 sources with both RM and redshift
data (the ``RM-redshift catalog''). We further derive a catalog of residual
rotation measure vs.\ redshift, in which we have subtracted the Galactic RM
contribution toward a subset of 3650 high-latitude sources (the
``RRM-redshift catalog'').  The resulting samples contain more than an
order of magnitude more sources than any previously published catalogs of RM
vs.\ redshift or RRM vs.\ redshift.

In our RRM-redshift catalog, we do not see any significant evidence that the variance of RRM
changes with redshift, in contrast to previous studies by \citet{welter} and
\citet{kronberg2008} who found that the RRMs of their sources showed
increased scatter at higher $z$. The overall standard deviation of the RRMs
in our catalog is 23~rad~m$^{-2}$, of which 13~rad~m$^{-2}$ is due to
errors associated with measurement and foreground removal,
$12-17$~rad~m$^{-2}$ is due to residual small-scale
fluctuations in the Galactic RM,
and $10-15$~rad~m$^{-2}$ is extragalactic
Faraday rotation that is not intrinsic to the emitting source and must
originate in intervening systems between the radio sources and the Milky Way.

We find a strong distinction between the fractional polarizations of radio
sources whose optical counterparts are galaxies and those whose optical
counterparts are quasars. The former can be highly polarized, representing
the extended lobes of radio galaxies, while the latter are only weakly
polarized, representing the radio cores of active galaxies.
Beyond this bimodality, we find no evolution of polarized fraction
with redshift for either the galaxy or quasar population considered
separately.

Finally, we identify a strong depolarization effect in our RRM-redshift
catalog, whereby sources with even modest residual Faraday rotation, $|{\rm RRM}| \ga
20$~rad~m$^{-2}$, have substantially reduced polarized fractions compared to
sources with RRMs near 0~rad~m$^{-2}$. We interpret this as beam
depolarization due to small-scale fluctuations in magnetic field strength
and gas density in the same intervening population that contributes to the
RRMs.  

A full consideration of the nature of the extragalactic source population
that produces the observed Faraday rotation and depolarization is beyond the
scope of this paper, but future investigations should consider the
relationship of RRM with $z$ for different sub-populations, and should study
the dependence of the observed depolarization on wavelength, observing
frequency and angular resolution.  Our analysis highlights the limitations
imposed by the contribution of the Galactic foreground RM, even in cases
where the foreground is modeled with a sophisticated algorithm that uses
more than 40\,000 RMs as input. A much denser RM grid, with each RM
determined far more robustly than from the two frequency channels of NVSS
data used here, is required to accurately account for the foreground
contribution and to identify subtle trends of polarization properties with
redshift.  Such a data set will be provided by the upcoming Polarization Sky
Survey of the Universe's Magnetism (POSSUM) on the Australian Square
Kilometre Array Pathfinder \citep{glt10}.  This next generation of
polarization data will be a powerful discriminant between different
mechanisms for extragalactic Faraday rotation measure and depolarization,
and can thus provide sensitive probes of magnetic field and electron density
as a function of cosmic time.

\acknowledgements 

We are indebted to Marion Schmitz for extraordinary levels of assistance in
using the NED database. We also thank Larry Rudnick, Paul Hancock, Helen
Johnston, John Ching, Elaine Sadler, Scott Croom and Julia Bryant for useful
discussions, and acknowledge valuable email correspondence with Ani Thakar,
Benjamin Alan Weaver, Harold Corwin, Landais Gilles, Marc Wenger, Francois
Ochsenbein, Amy Kimball, Jim Condon and Matthew Colless.  B.~M.~G. and T.~R.
acknowledge the support of the Australian Research Council through grants
FF0561298, FL100100114 and FS100100033.  This research has made use of: the
SIMBAD database, operated at Centre de Donn\'{e}es astronomiques de
Strasbourg, France; the NASA/IPAC Extragalactic Database (NED) which is
operated by the Jet Propulsion Laboratory, California Institute of
Technology, under contract with NASA; NASA's Astrophysics Data System
Abstract Service; the SDSS, for which funding has been provided by the
Alfred P. Sloan Foundation, the Participating Institutions, the National
Science Foundation, and the U.S. Department of Energy Office of Science; the
2dF/6dF QSO Redshift Surveys (2QZ/6QZ), which were compiled by the 2QZ/6QZ
teams from observations made with the 2-degree Field and 6-degree Field on
the Anglo-Australian Telescope; and the IDL Astronomy User's Library at
Goddard Space Flight Center (available at http://idlastro.gsfc.nasa.gov/).
Some of the results in this paper have been derived using the HEALPix
\citep{ghb+05} package.  The National Radio Astronomy Observatory is a
facility of the National Science Foundation operated under cooperative
agreement by Associated Universities, Inc.

Facilities:
\facility{AAT}
\facility{Sloan}
\facility{UKST}
\facility{VLA}

\bibliography{catalogue_paper}

\clearpage

\begin{deluxetable}{ll}
\tablecaption{Association classes between NVSS radio sources of \citet{taylor}
and their optical counterparts.\label{tab:cat_classes}  }
\tabletypesize{\small}
\tablewidth{0pt}  
\tablehead{\colhead{Class}\hspace{4ex} & \colhead{Description}}
\startdata
  A\dotfill & Match using FIRST  \\
  ~~A(i)\dotfill & Close core match with FIRST  \\
  ~~A(ii)\dotfill & Core match with FIRST  \\
  ~~A(iii)\dotfill & Single NVSS; closely matched double in FIRST  \\
  ~~A(iv)\dotfill & Single NVSS; double in FIRST  \\
  ~~A(v)\dotfill & Single NVSS; closely matched triple in FIRST  \\
  ~~A(vi)\dotfill & Single NVSS; triple in FIRST  \\
  ~~A(vii)\dotfill & Double-lobed in NVSS with FIRST  \\
  B\dotfill & Core match in NVSS; either no FIRST detection or no FIRST data \\
  C\dotfill & Manual visual identification \\
  D\dotfill & Association made in NED \\
  E\dotfill & Association made in SIMBAD 
\enddata
\end{deluxetable}

\begin{deluxetable}{cc}
\tablecaption{\label{tab:cat} Prioritization of surveys and databases
in selecting a final redshift, in cases where a radio source
from \citet{taylor} has matches in more than one optical
redshift catalog.}
\tabletypesize{\small}
\tablewidth{0pt}
\tablehead{\colhead{Priority} & \colhead{Survey or Database}}
\startdata
1 & SDSS \\
2 & 6dFGS \\
3 & 2dFGRS \\
4 & 2QZ/6QZ \\
5 & NED \\
6 & SIMBAD 
\enddata
\end{deluxetable}

\input{rm_z_catalog}

\input{table4}

\input{rrm_z_catalog}

\begin{figure}
\centerline{\psfig{file=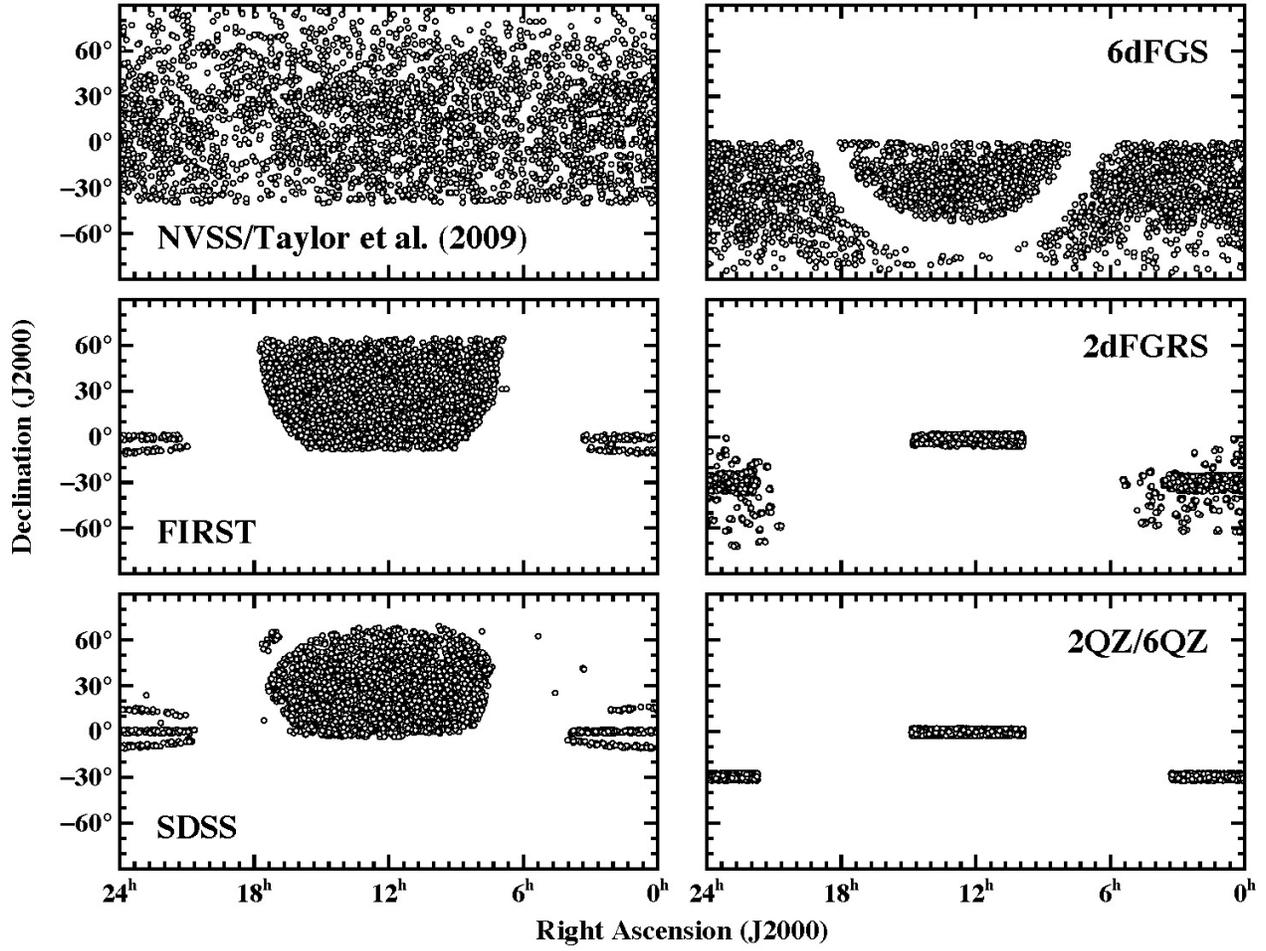,width=\textwidth}}
\caption{The sky coverage for each of the surveys used in the creation of
our RM vs.\ redshift catalog, shown using sparse sampling.\label{fig:sky
coverage} }
\end{figure}

\begin{figure}
\centerline{\psfig{file=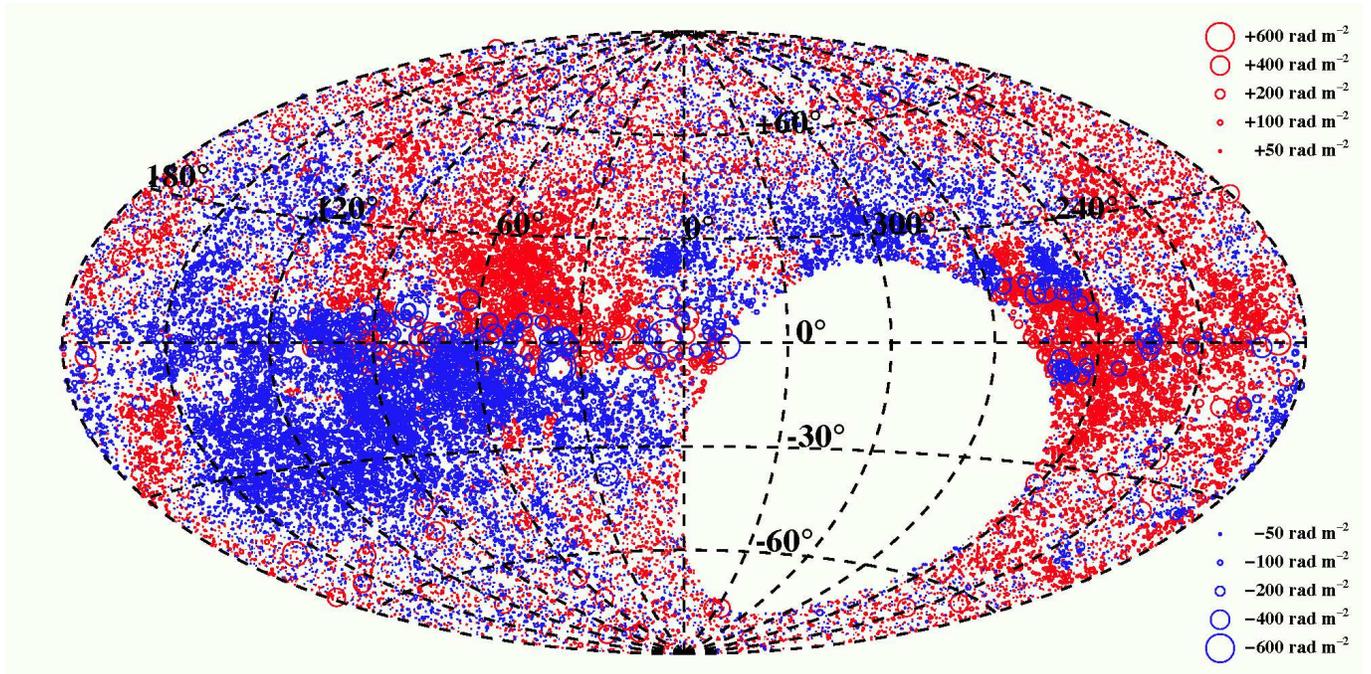,width=\textwidth}}
\caption{RMs of 37\,543 extragalactic sources over the sky above a
declination of $-40^{\circ}$ from the survey by \citet{taylor}.  Red circles
correspond to positive RMs while blue circles are negative.  The size of the
circle scales linearly with the magnitude of the RM.  The sources are mapped
onto a Galactic co-ordinate system.\label{fig:taylor_cat}}
\end{figure}

\begin{figure}
\centerline{\psfig{file=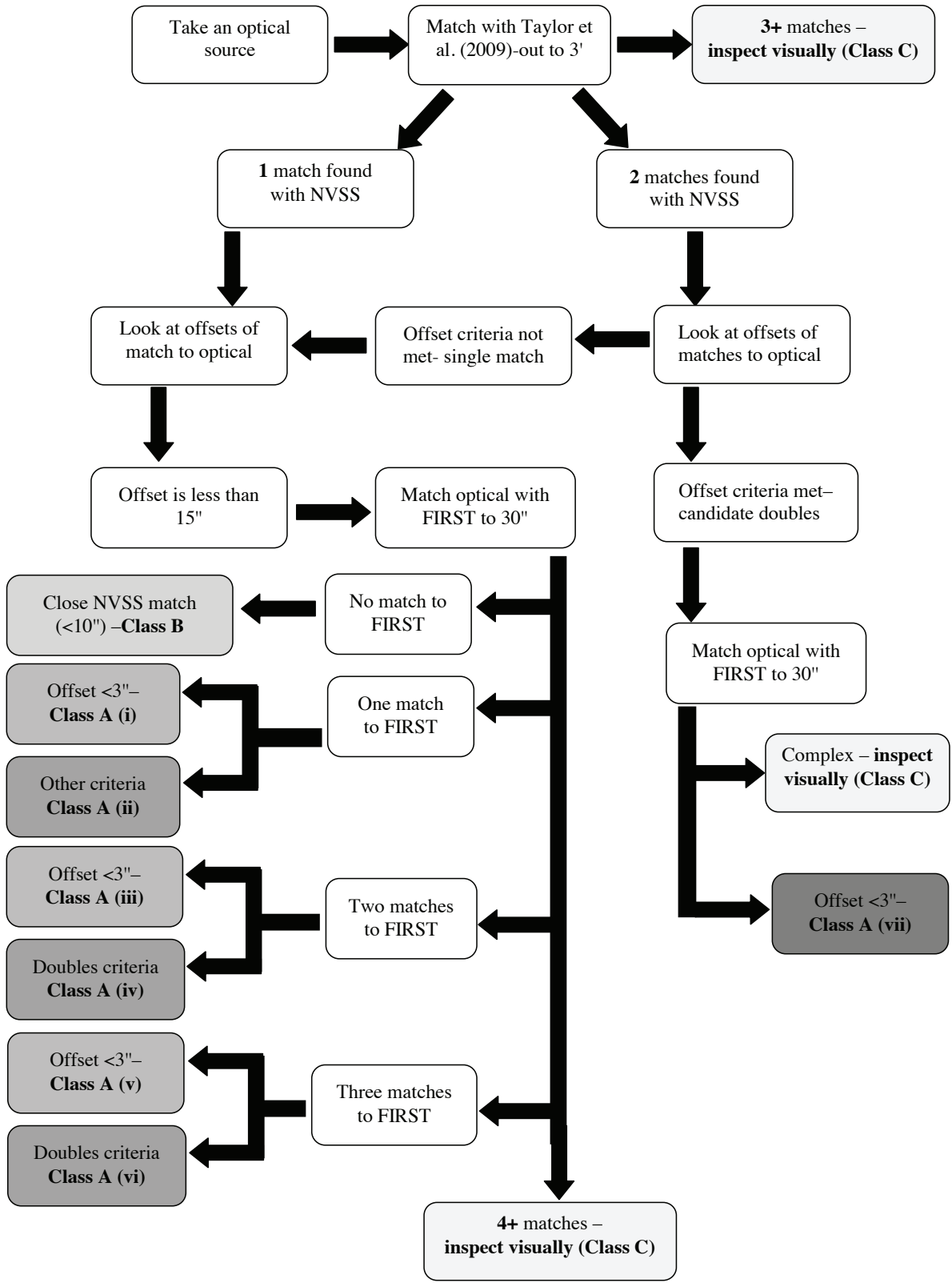,width=0.9\textwidth}}
\caption{A simplified flowchart representation of the algorithm implemented
in this paper to make associations between optical and radio sources in
various surveys. Final association classes are shown shaded, with the same
shading used for classes with similar admission criteria. 
\label{fig:algorithm}}
\end{figure}

\begin{figure} 
\centerline{\psfig{file=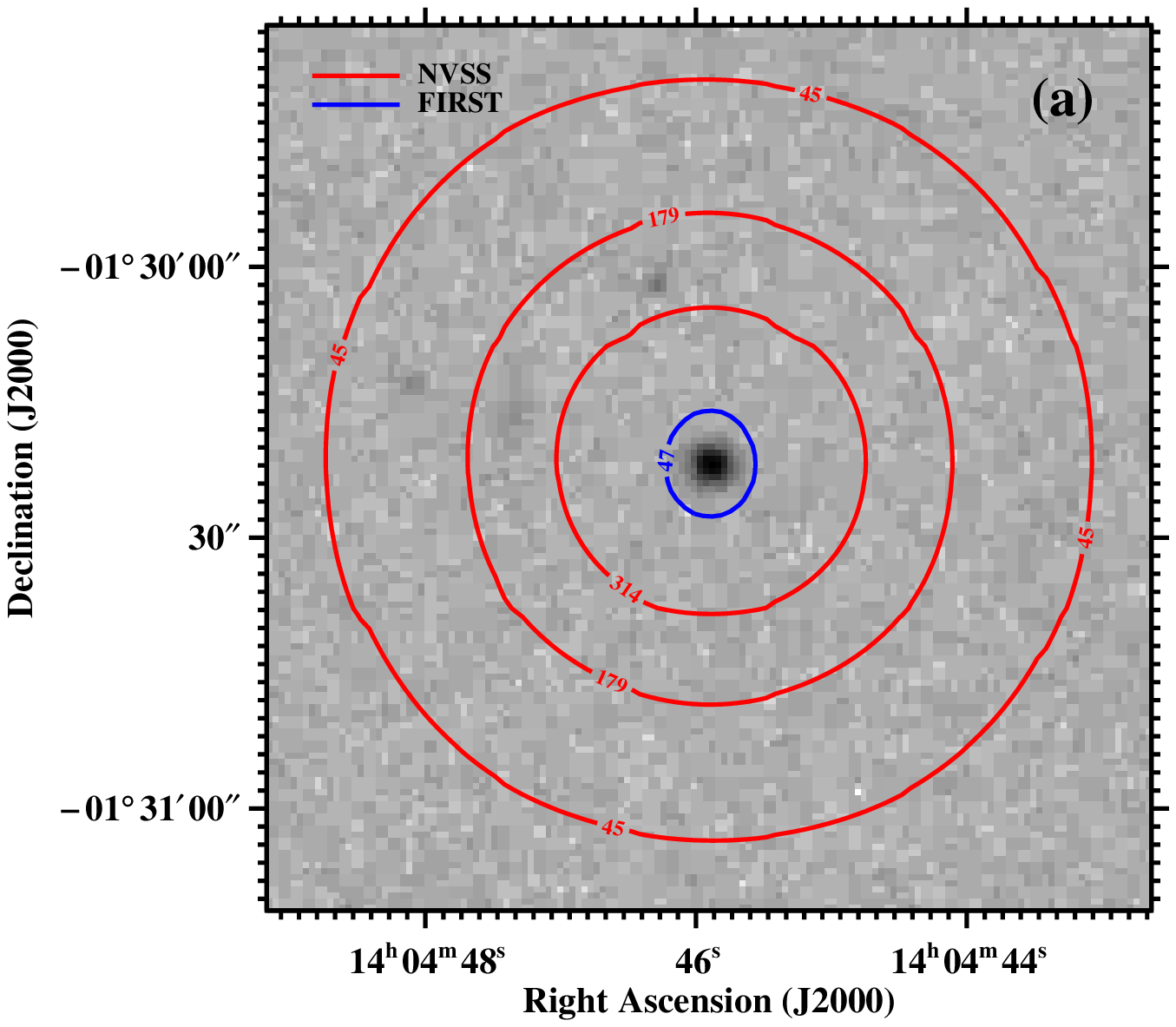,width=0.5\textwidth}
            \psfig{file=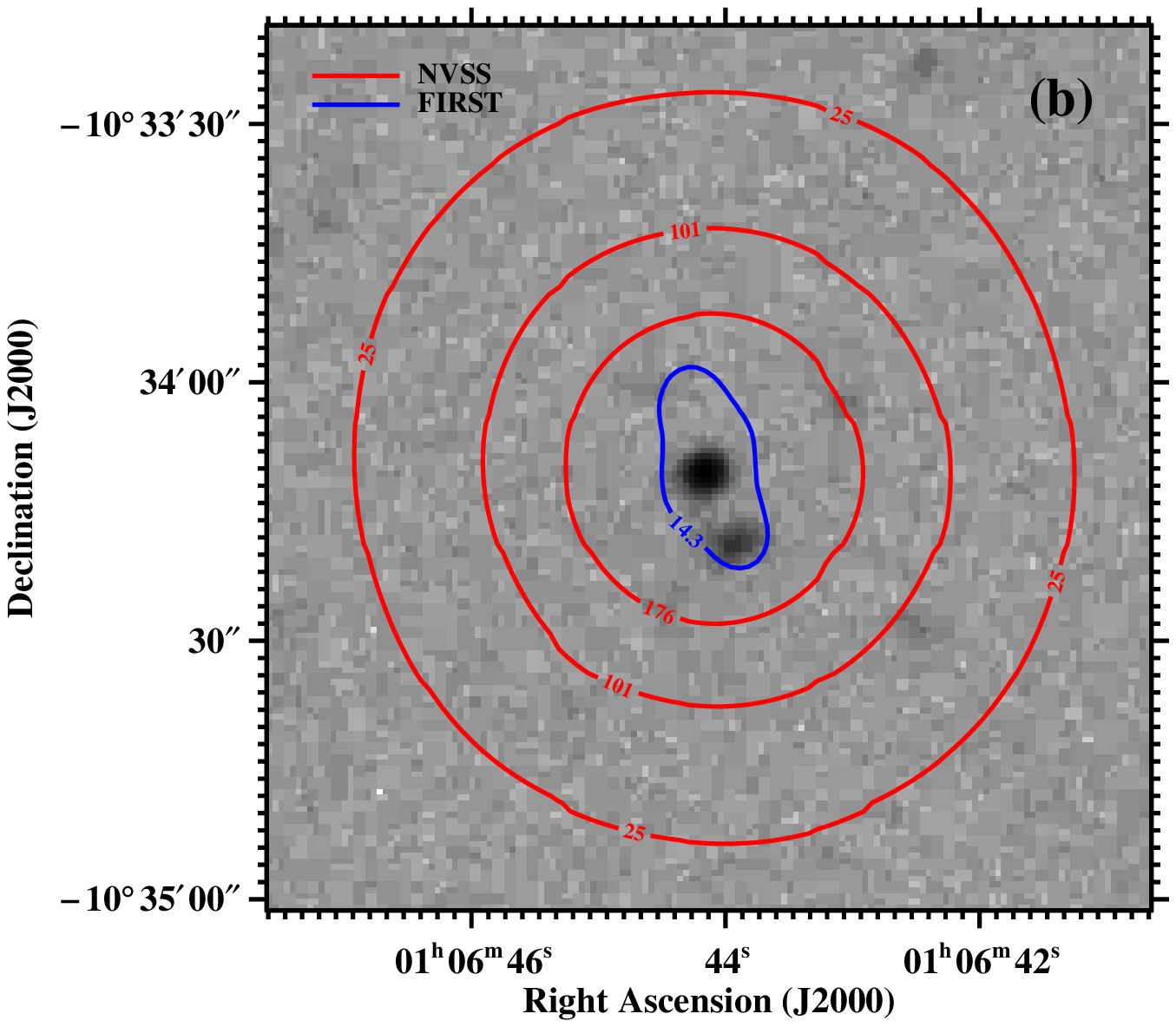,width=0.5\textwidth}}
\caption{Examples of class A(i) associations (close core match with FIRST
data) made with our algorithm. Optical data are shown in gray-scale with
NVSS radio contours overlaid in red and FIRST radio contours overlaid in
blue. (a) NVSS J140445$-$013021: class A(i) association for both SDSS and
2QZ/6QZ. (b) NVSS J010644$-$103409: class A(i) association for both SDSS and
6dFGS.  All contour levels are measured in mJy
beam$^{-1}$. In this and subsequent related Figures, optical
data are taken from SDSS where available, and otherwise
from SuperCOSMOS.\label{fig:example_associations1}}
\end{figure}

\begin{figure} 
\centerline{\psfig{file=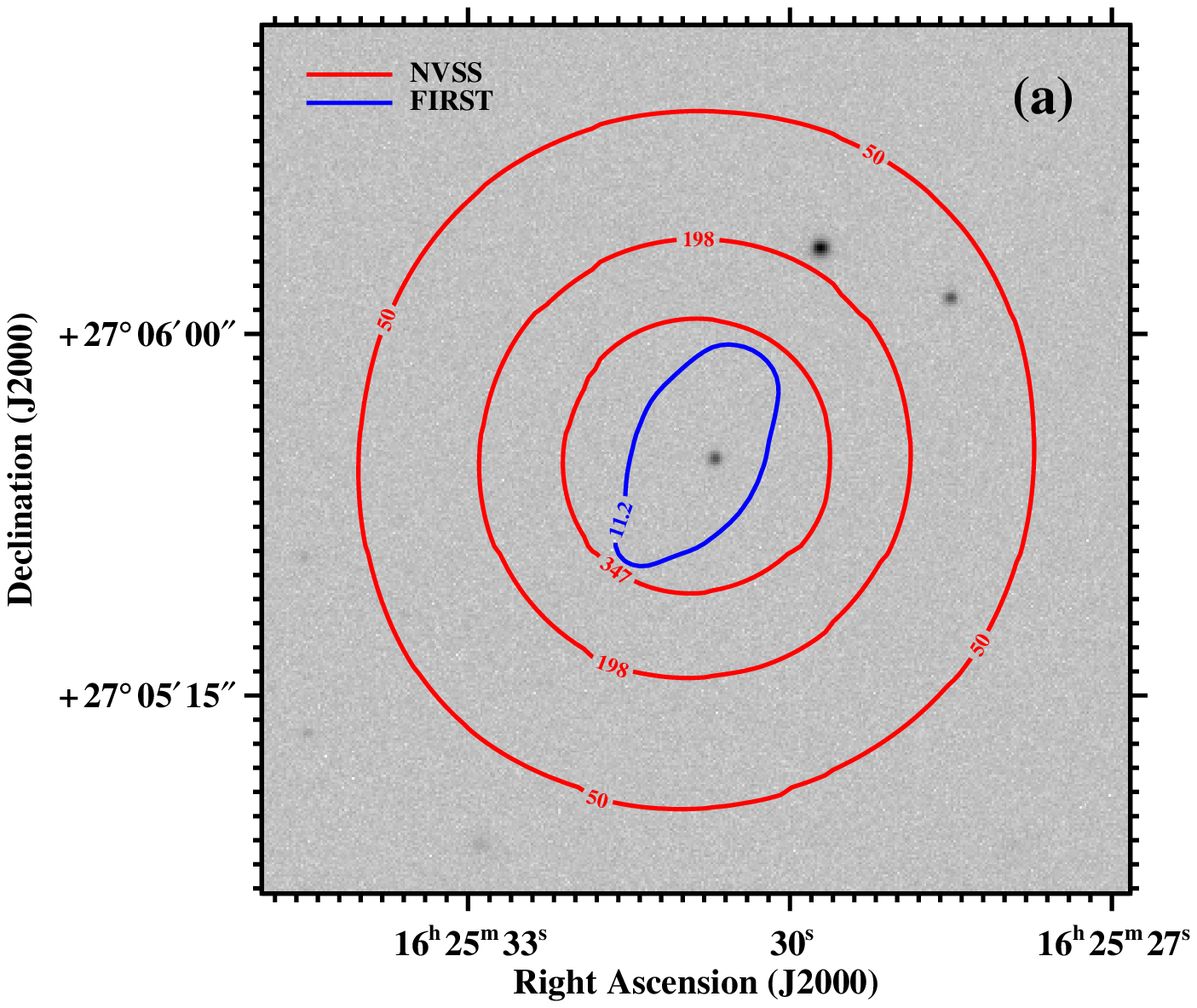,width=0.5\textwidth}
            \psfig{file=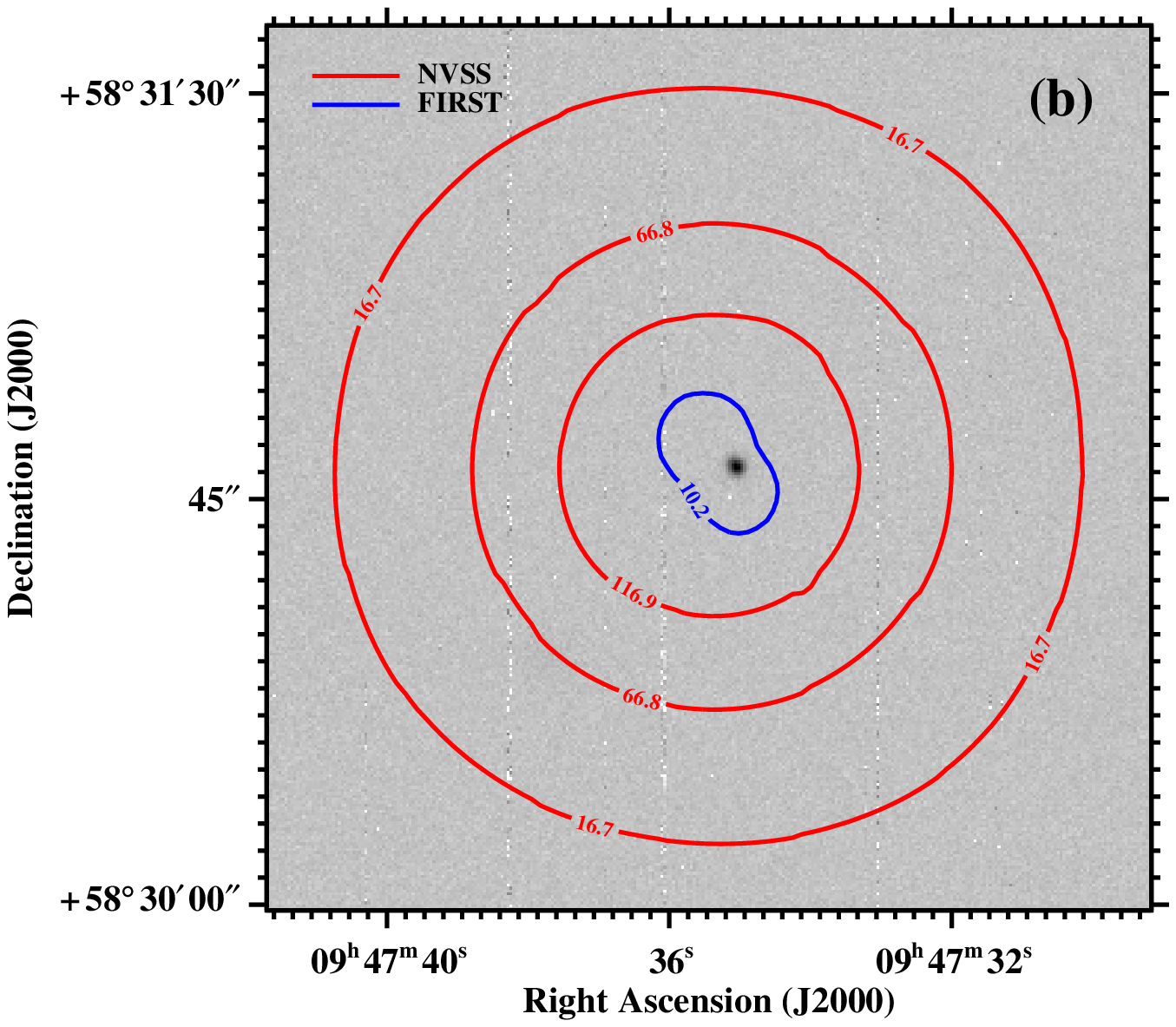,width=0.5\textwidth}} 
\caption{As for Figure~\ref{fig:example_associations1}, but showing examples
of class A(ii) associations (core match with FIRST data).  (a) NVSS
J162530$+$270544: class A(ii) association with the SDSS. (b) NVSS
J094735$+$583048: class A(ii) association with the SDSS.
\label{fig:example_associations2}}
\end{figure}

\begin{figure} 
\centerline{\psfig{file=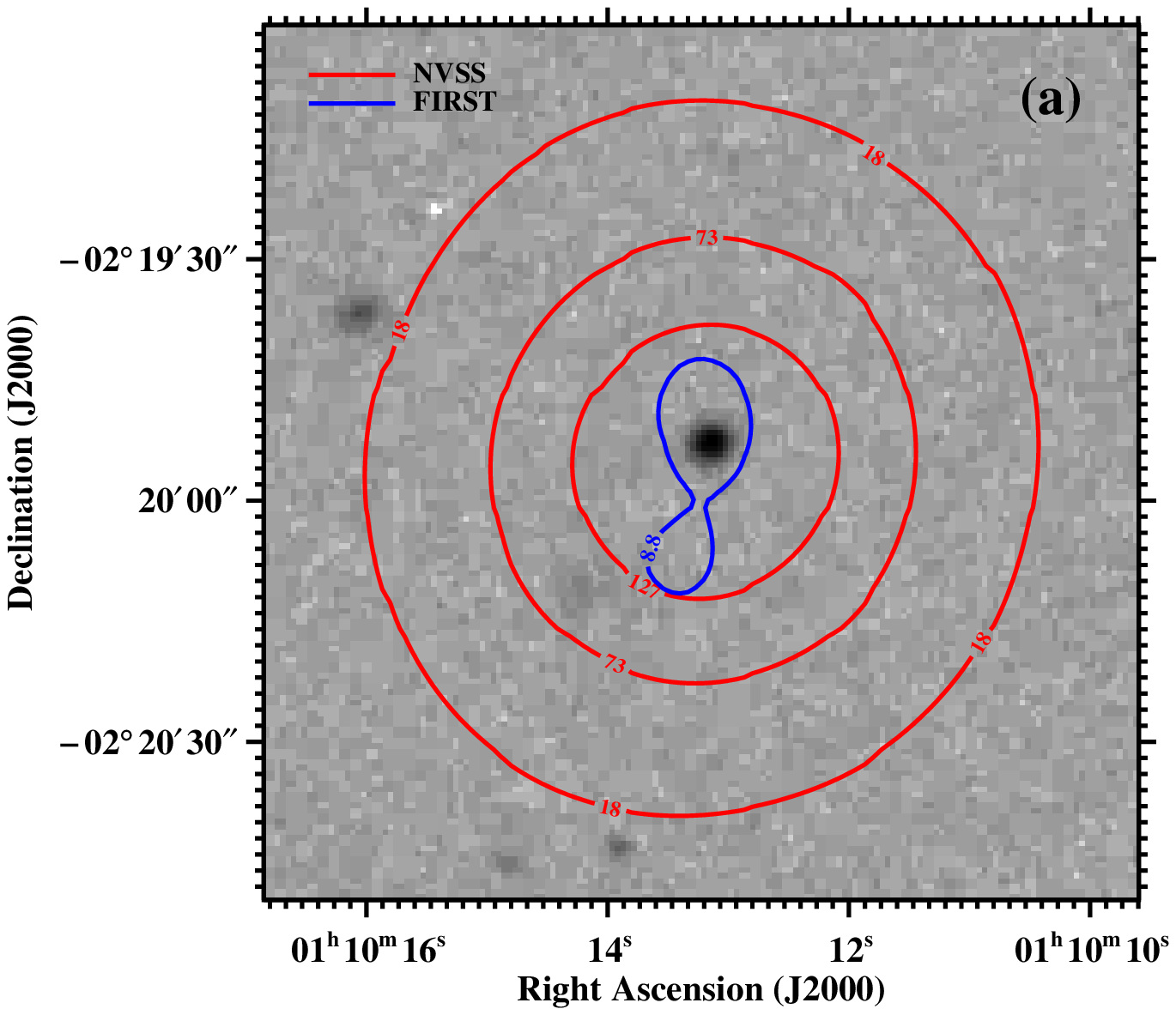,width=0.5\textwidth}
            \psfig{file=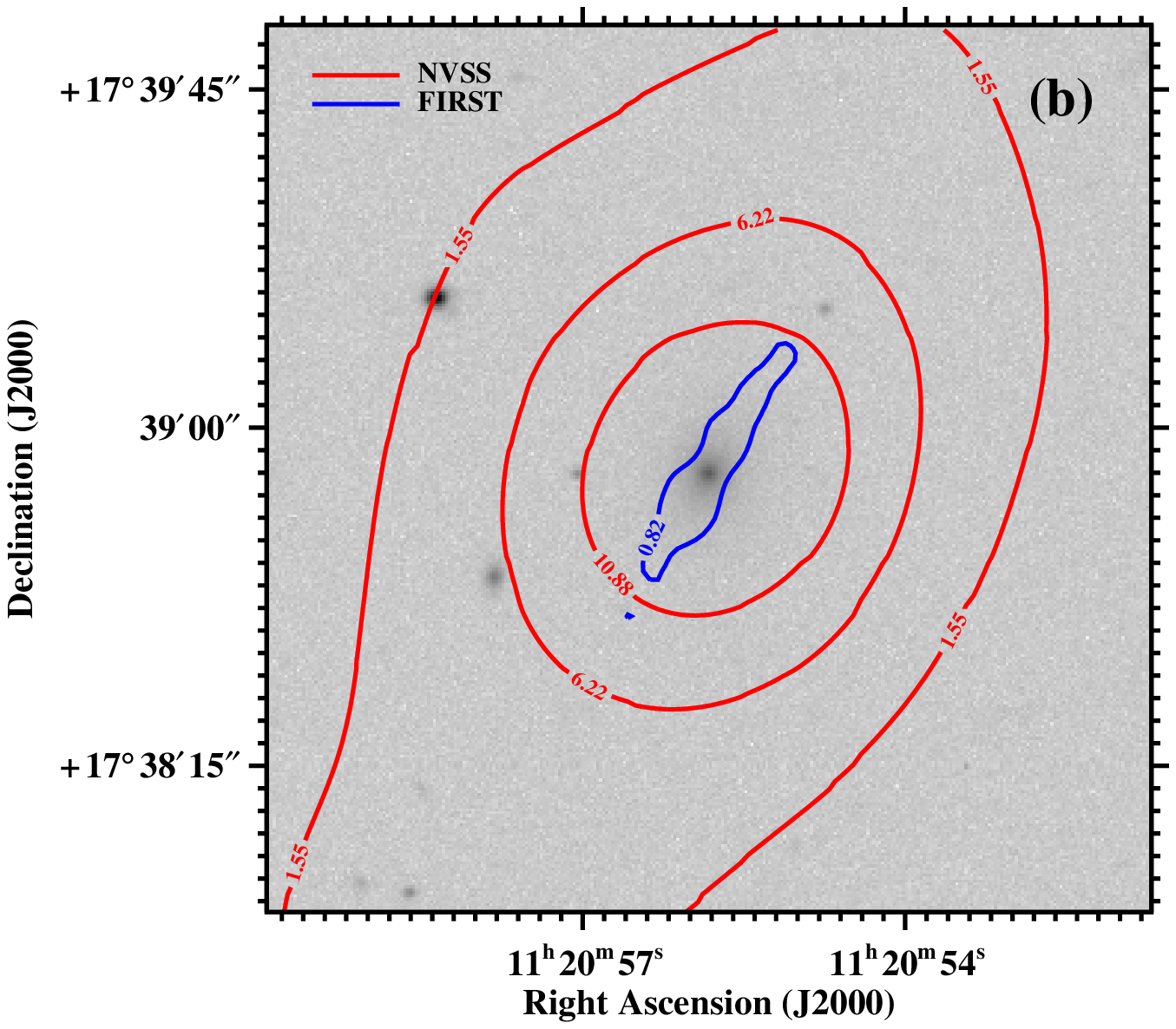,width=0.5\textwidth}} 
\caption{As for Figure~\ref{fig:example_associations1}, but showing examples
of class A(iii) associations (single NVSS source, closely matched double
source seen with FIRST). (a) NVSS J011013$-$021954: class A(iii) association
with the 6dFGS.  (b) NVSS J112055$+$173854: class A(iii) association with
the SDSS. 
\label{fig:example_associations3}}
\end{figure}

\begin{figure} 
\centerline{\psfig{file=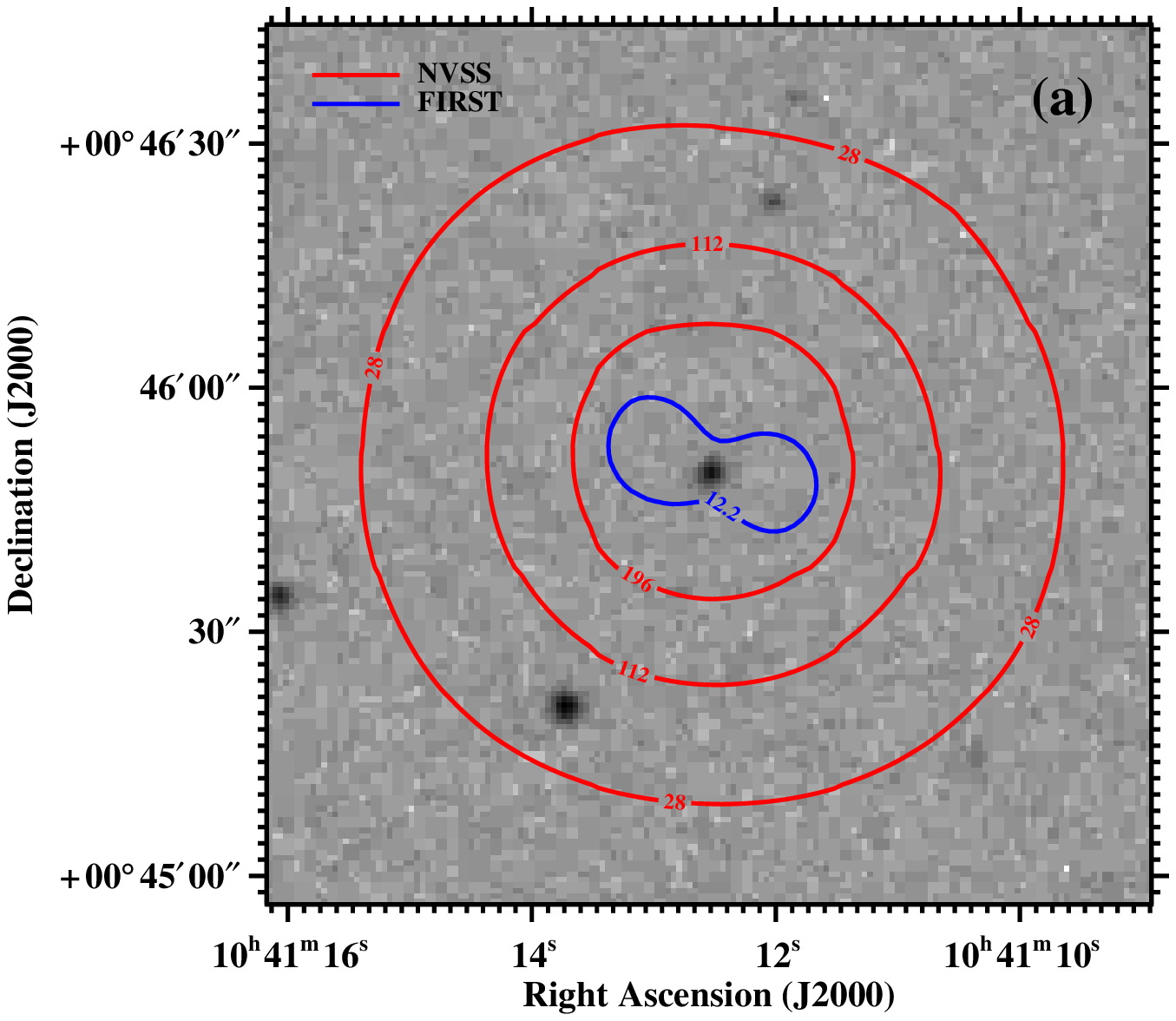,width=0.5\textwidth}
            \psfig{file=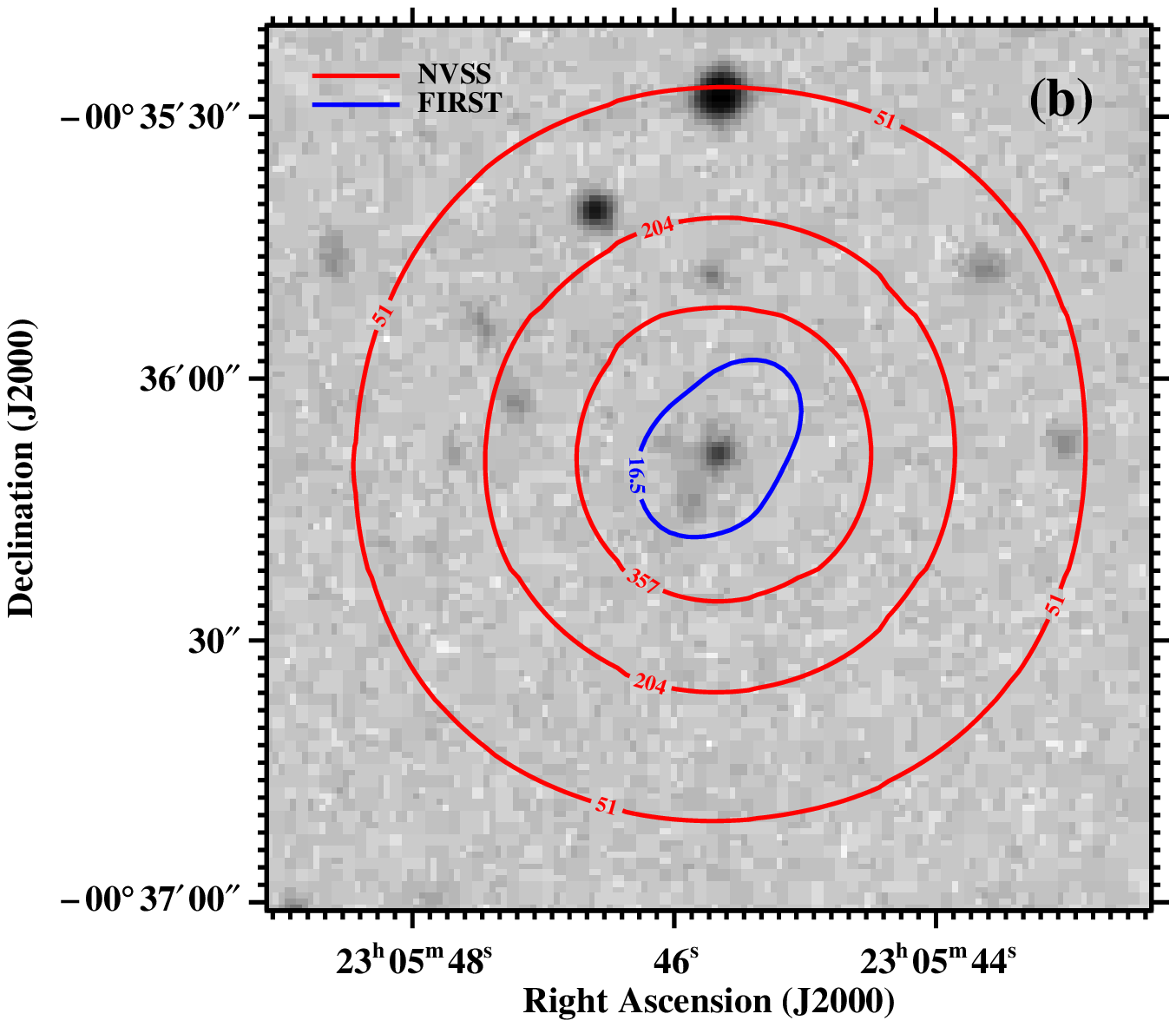,width=0.5\textwidth}} 

\caption{As for Figure~\ref{fig:example_associations1}, but showing examples
of class A(iv) associations (single NVSS source, double source seen with
FIRST).  (a) NVSS J104112$+$004550: class A(iv) association with the
2QZ/6QZ. (b) NVSS J230545$-$003608: class A(iv) association with the SDSS
and 2dFGRS.
\label{fig:example_associations4}}
\end{figure}

\begin{figure} 
\centerline{\psfig{file=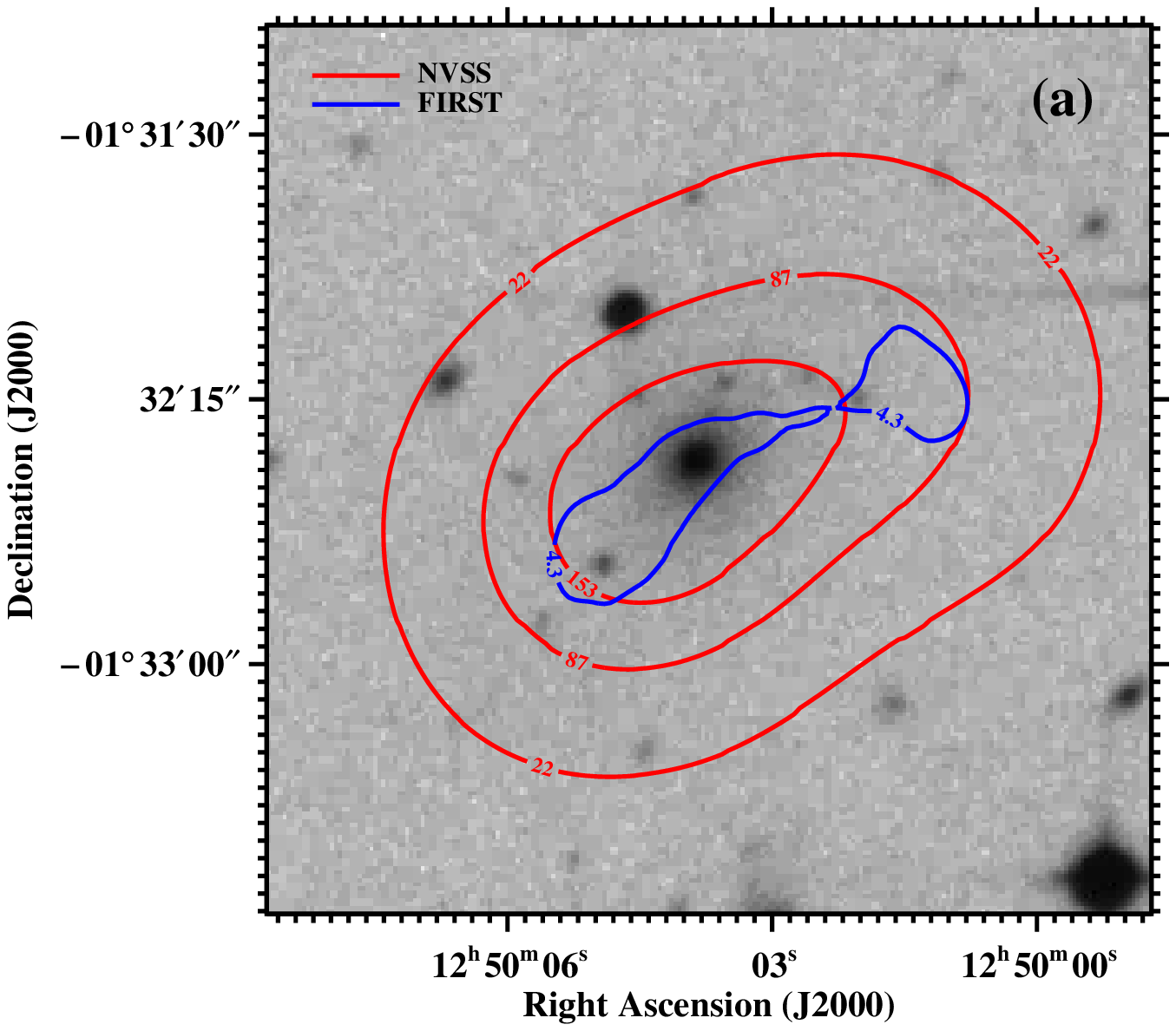,width=0.5\textwidth}
            \psfig{file=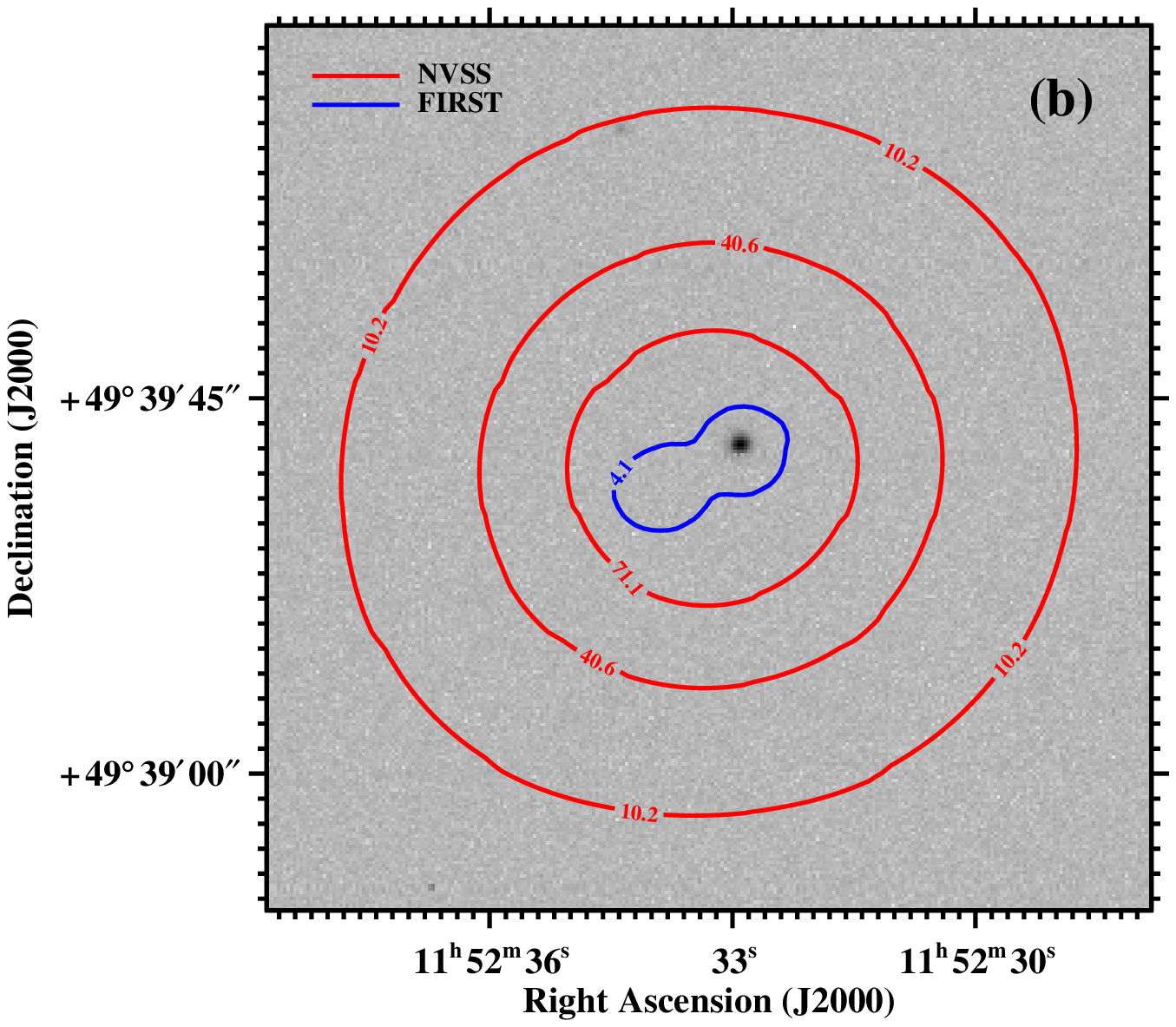,width=0.5\textwidth}} 
\caption{As for Figure~\ref{fig:example_associations1}, but showing examples
of class A(v) associations (single NVSS source, closely matched triple
source seen with FIRST).  (a) NVSS J125003$-$013226 : class A(v) association
with the 2dFGRS. (b) NVSS J115233$+$493937: class A(v) association with the
SDSS.
\label{fig:example_associations5}}
\end{figure}

\begin{figure} 
\centerline{\psfig{file=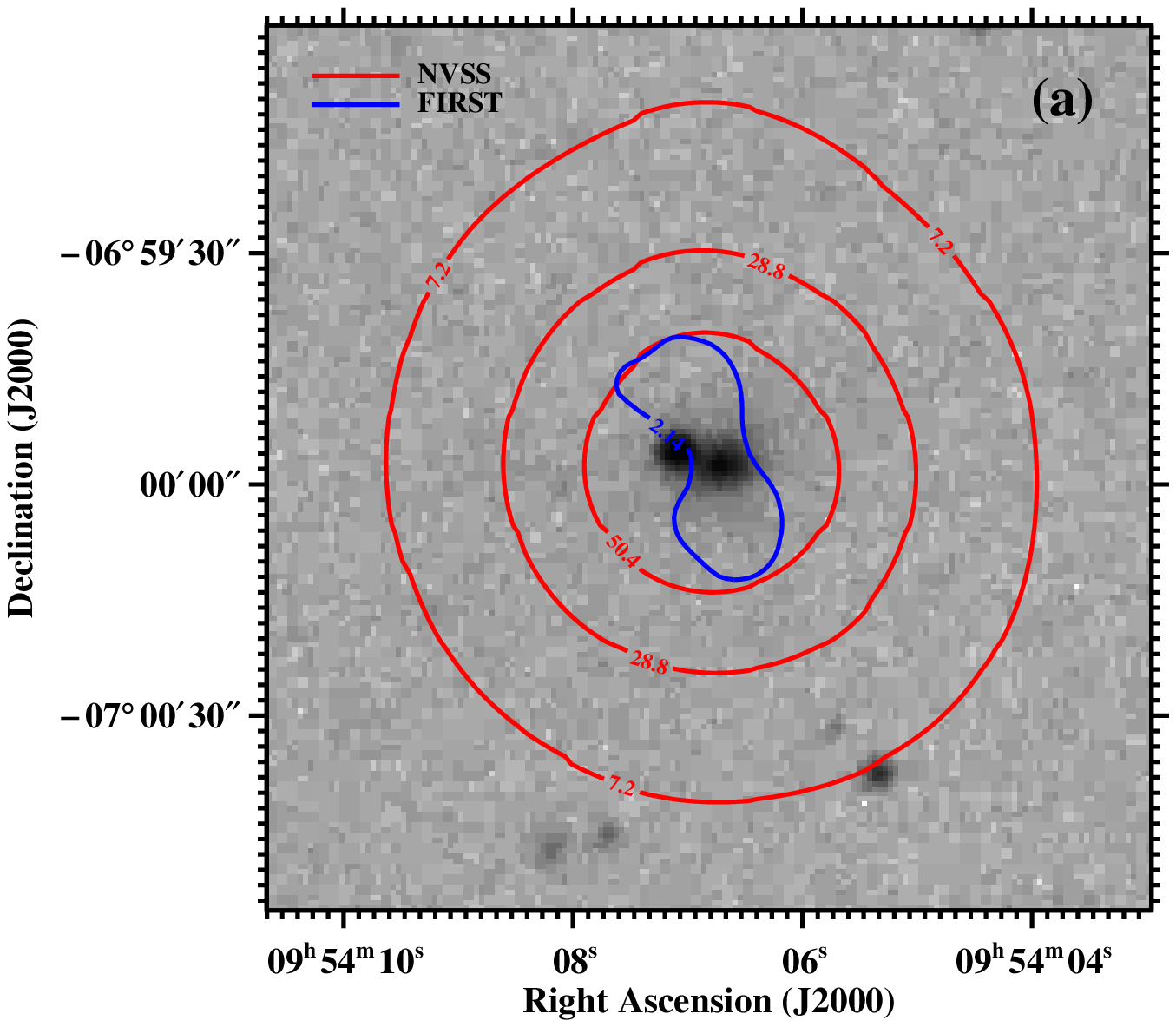,width=0.5\textwidth}
            \psfig{file=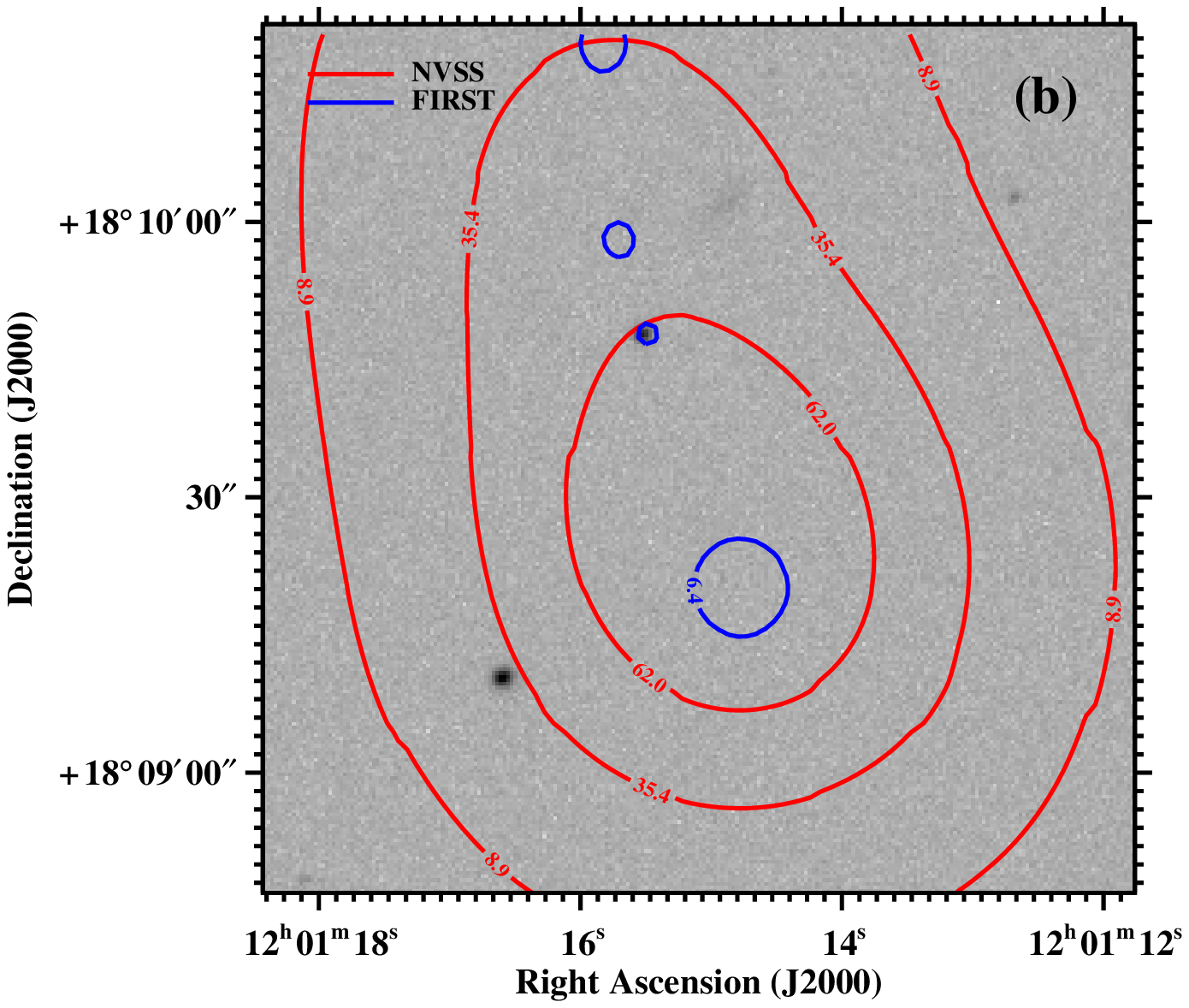,width=0.5\textwidth}} 
\caption{As for Figure~\ref{fig:example_associations1}, but showing examples
of class A(vi) associations (single NVSS source, triple source seen with
FIRST).  (a) NVSS J095406$-$065957: class A(vi) association with the 6dFGS.
(b) NVSS J120115$+$180934: class A(vi) association with the SDSS; here, the
faint optical source lies within the second FIRST component from the bottom.
\label{fig:example_associations6}}
\end{figure}

\begin{figure} 
\centerline{\psfig{file=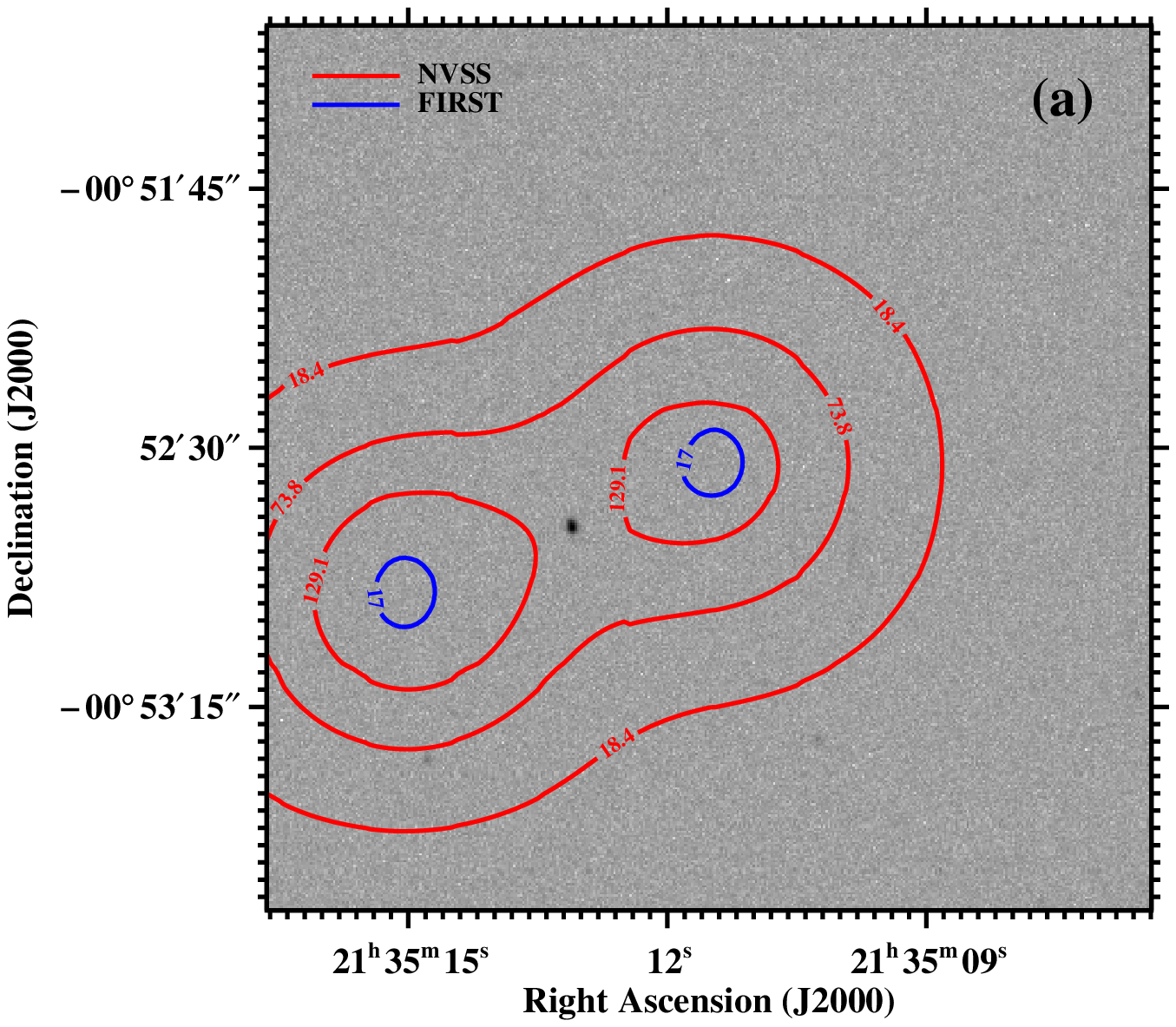,width=0.5\textwidth}
            \psfig{file=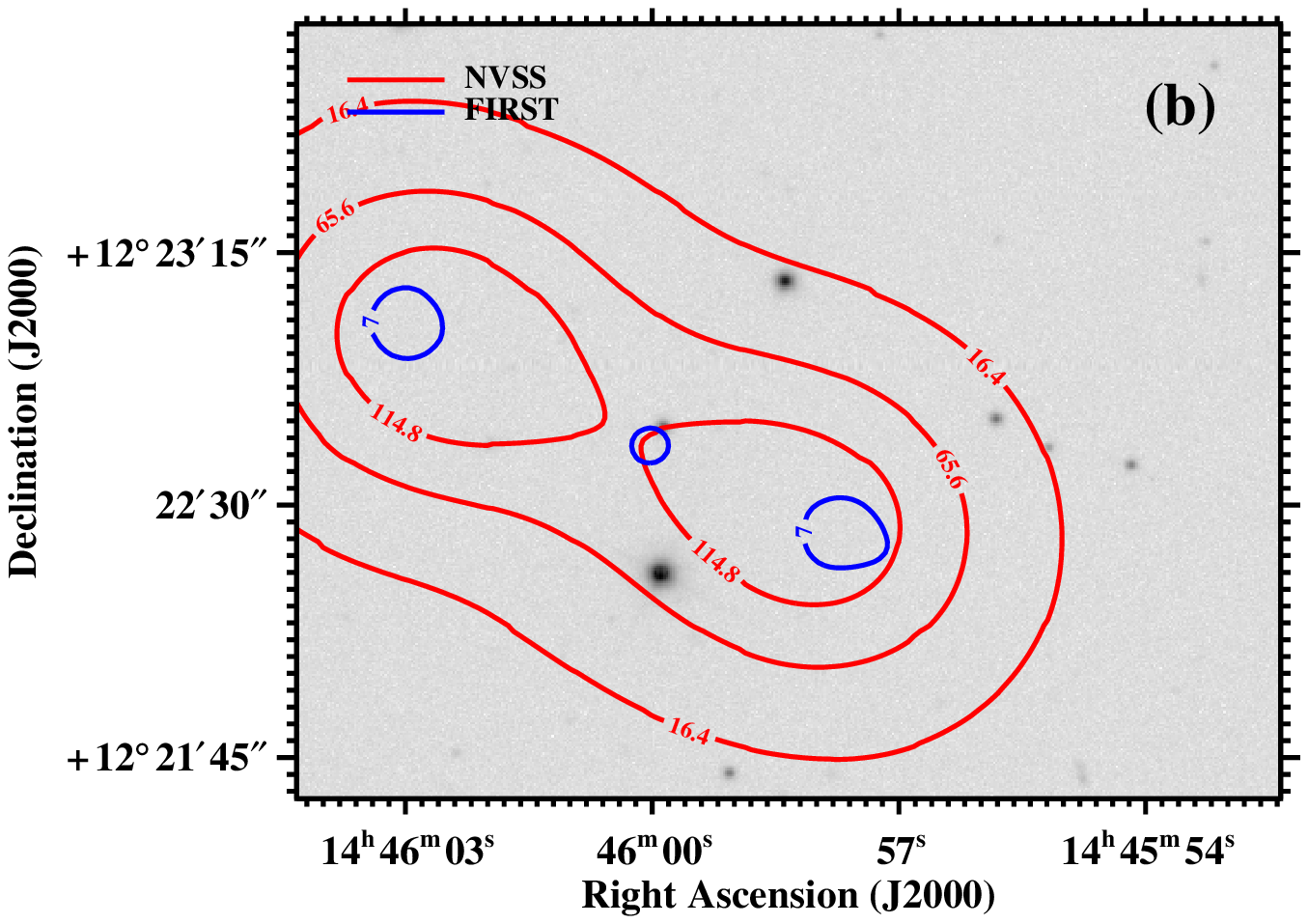,width=0.5\textwidth}} 
\caption{As for Figure~\ref{fig:example_associations1}, but showing examples
of class A(vii) associations (double-lobed radio source in NVSS, also
detected in FIRST).  (a) NVSS J213515$-$005255 and NVSS J213511$-$005233:
class A(vii) association with the SDSS; the optical source lies in the
middle of the FIRST contours. (b) NVSS J144558$+$122228 and NVSS
J144602+122258: class A(vii) association with the SDSS. 
\label{fig:example_associations7}}
\end{figure}

\begin{figure} 
\centerline{\psfig{file=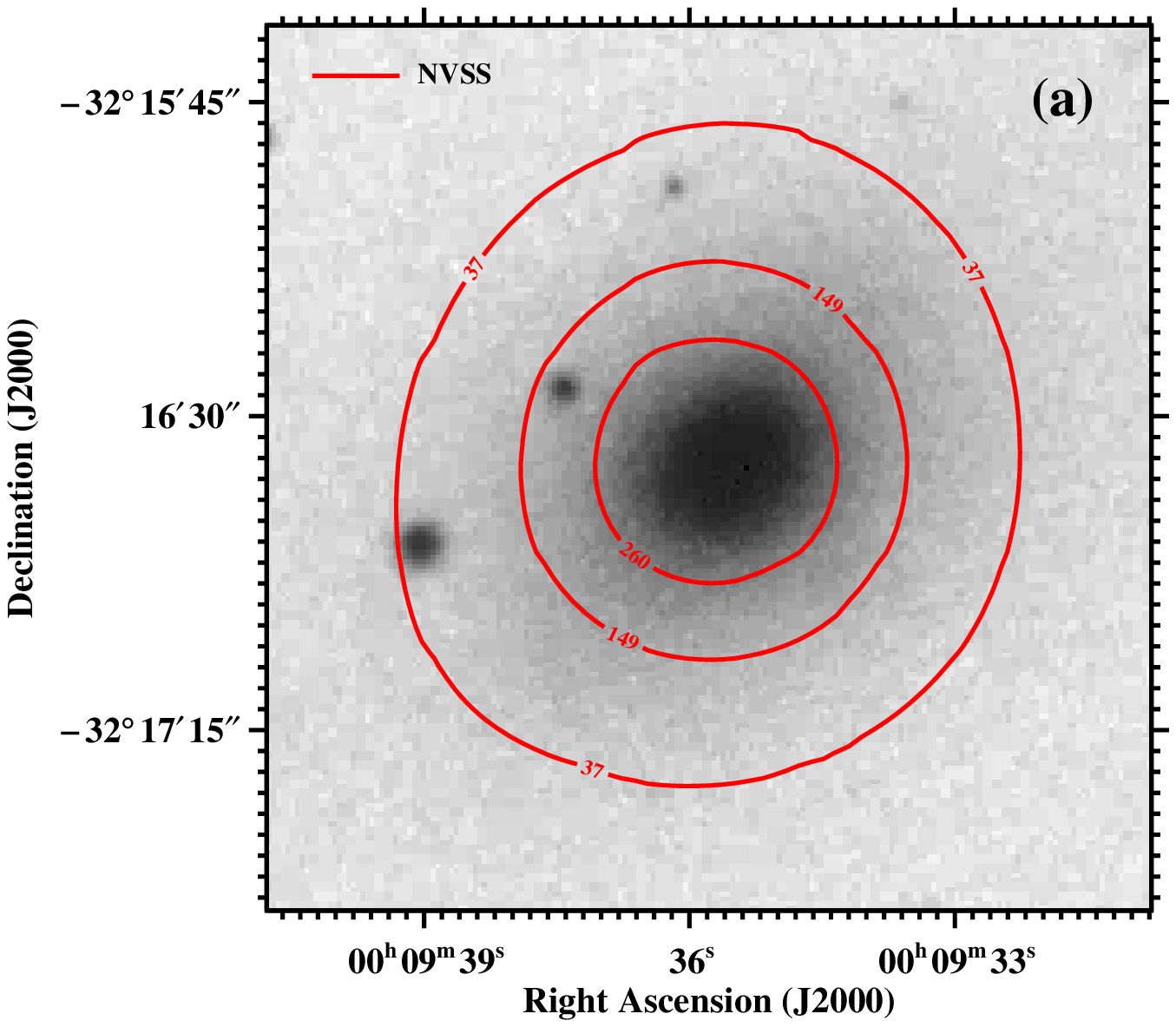,width=0.5\textwidth}
            \psfig{file=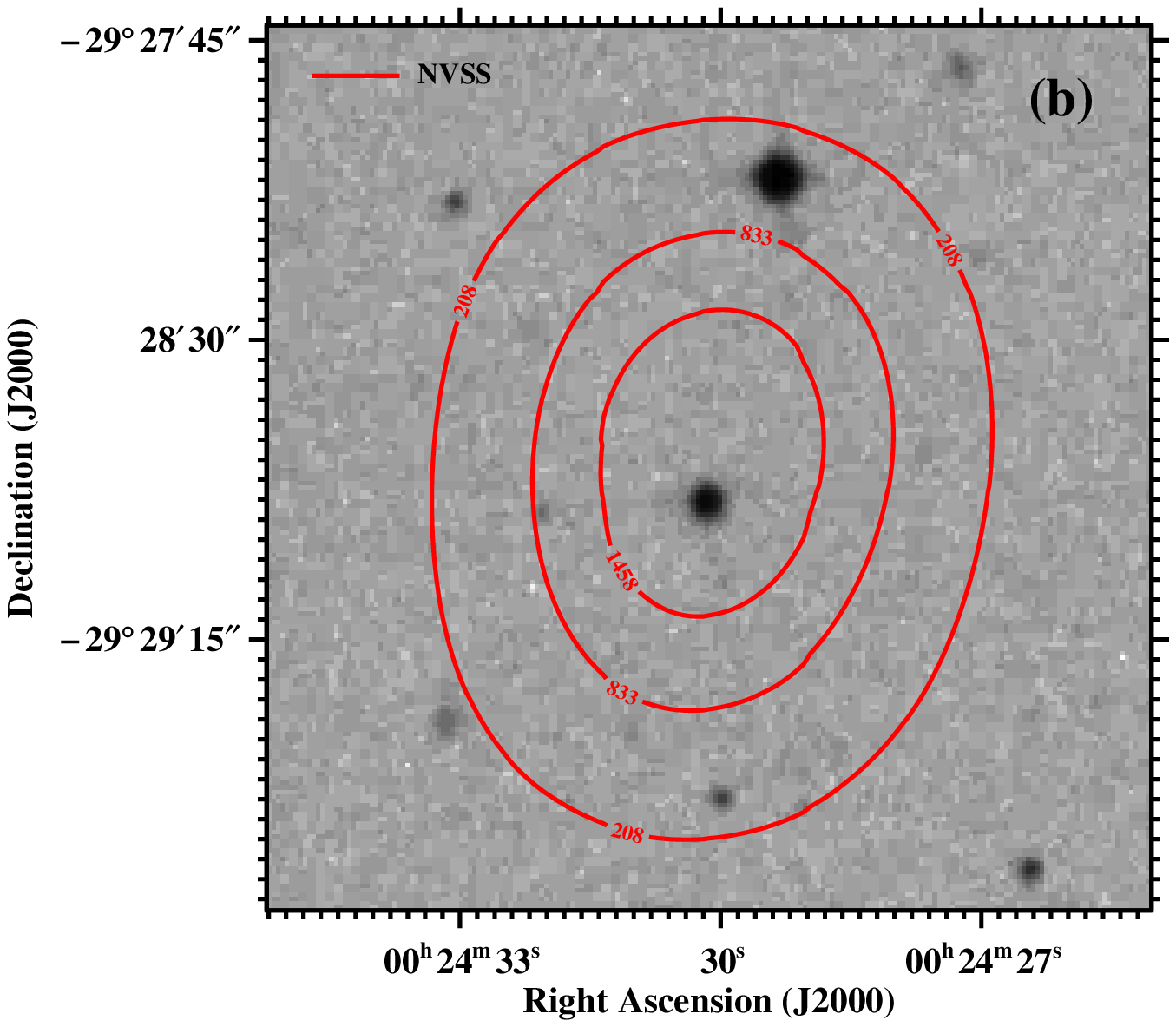,width=0.5\textwidth}} 
\caption{As for Figure~\ref{fig:example_associations1}, but showing examples
of class B associations (core match with NVSS but either no FIRST detection
or no FIRST observations).  (a) NVSS J000935$-$321636: class B association
for both 6dFGS and 2dFGRS. (b) NVSS J002430$-$292848: class B association
with the 2QZ/6QZ; in this case, the optical source is slightly offset from
the radio center. 
\label{fig:example_associations8}}
\end{figure}

\begin{figure} 
\centerline{\psfig{file=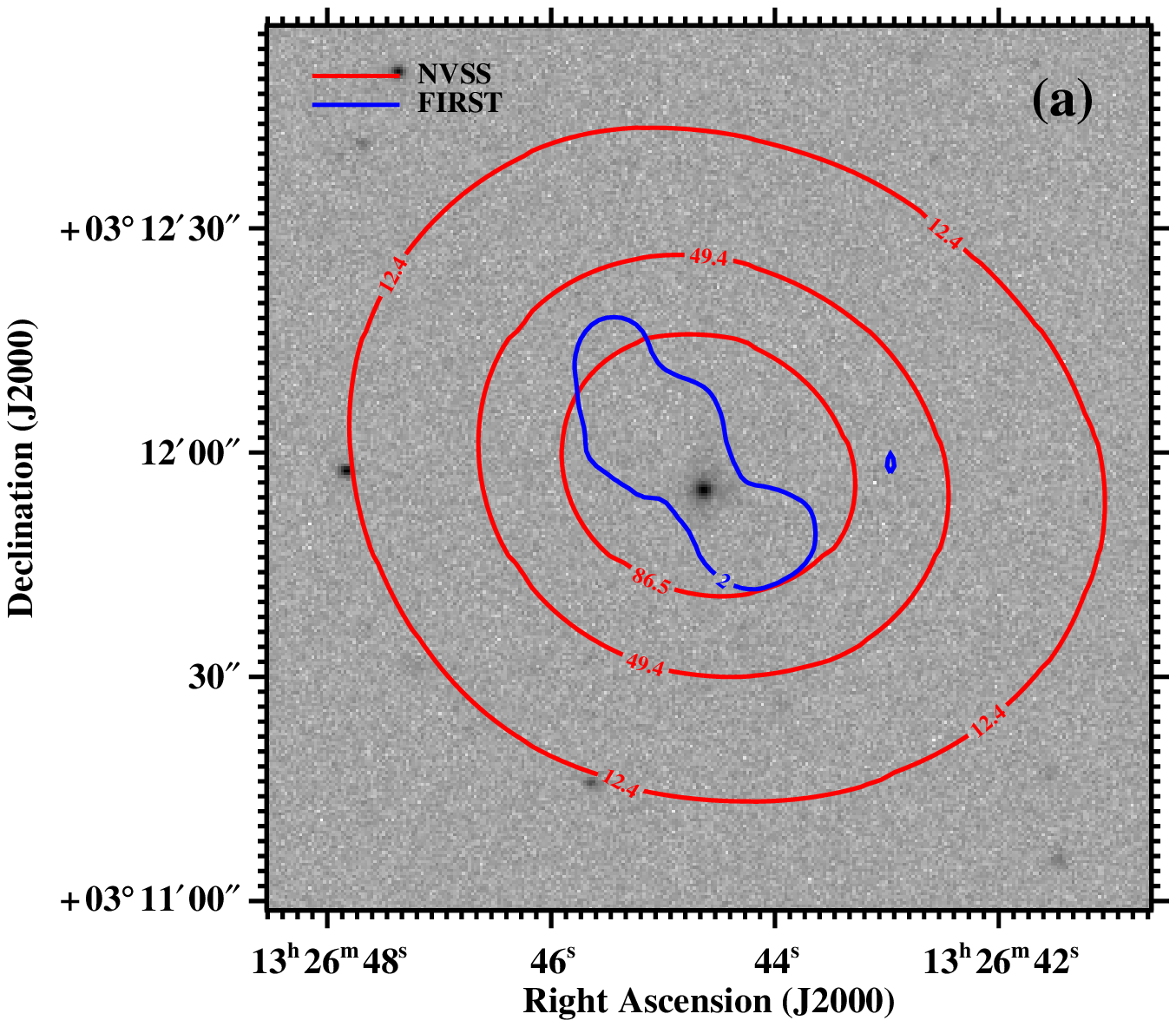,width=0.5\textwidth}
            \psfig{file=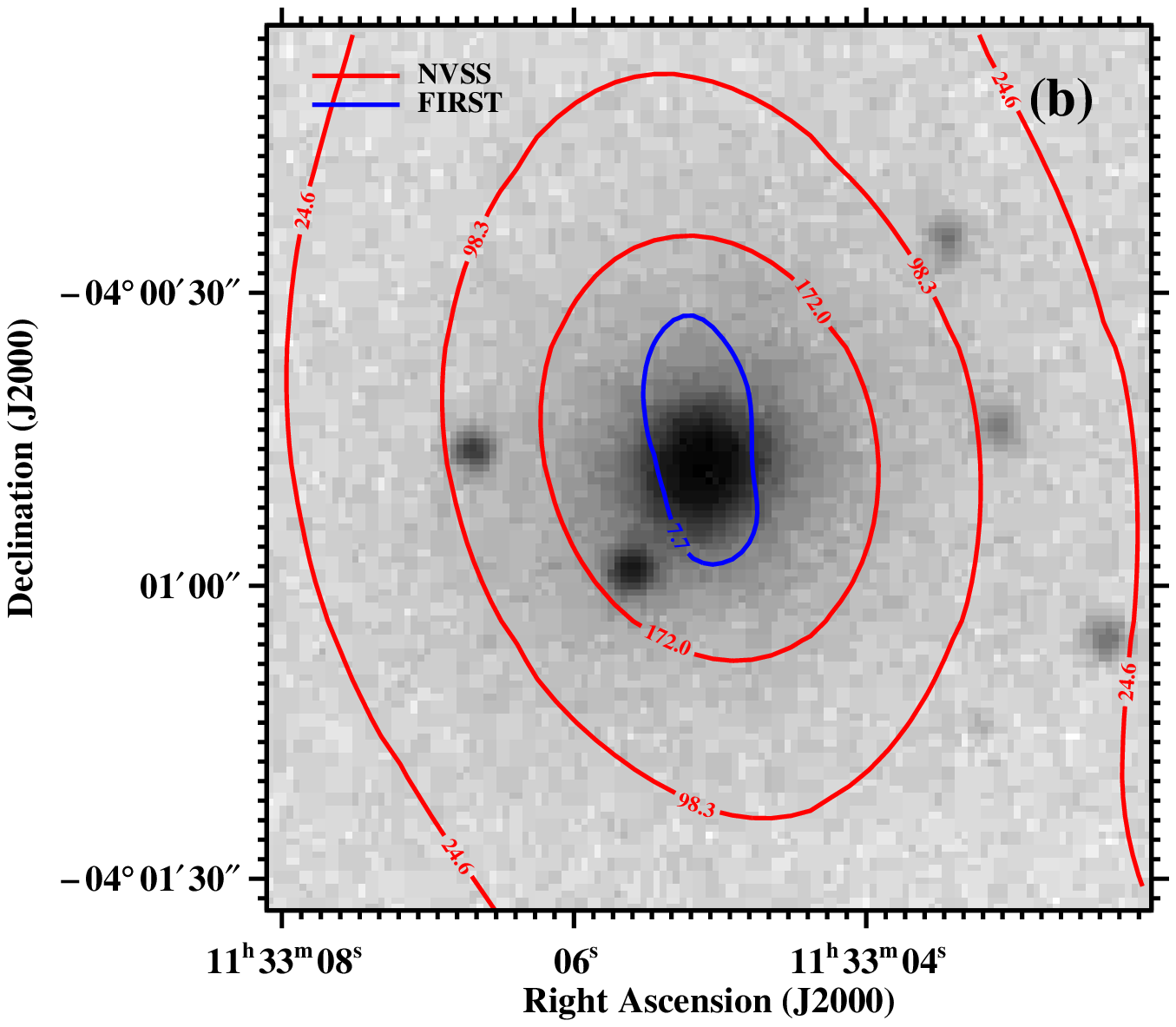,width=0.5\textwidth}} 
\caption{As for Figure~\ref{fig:example_associations1}, but showing examples
of class C associations (complex sources requiring manual visual
inspection).  (a) NVSS J132644$+$031158: class C association with the SDSS;
here the complexity is the presence of four FIRST components. (b) NVSS
J113305$-$040047: class C association with the 6dFGS; the complexity here is
the extremely extended nature of the NVSS source, which FIRST does not
detect.  
\label{fig:example_associations9}}
\end{figure}

\begin{figure}
\centerline{\psfig{file=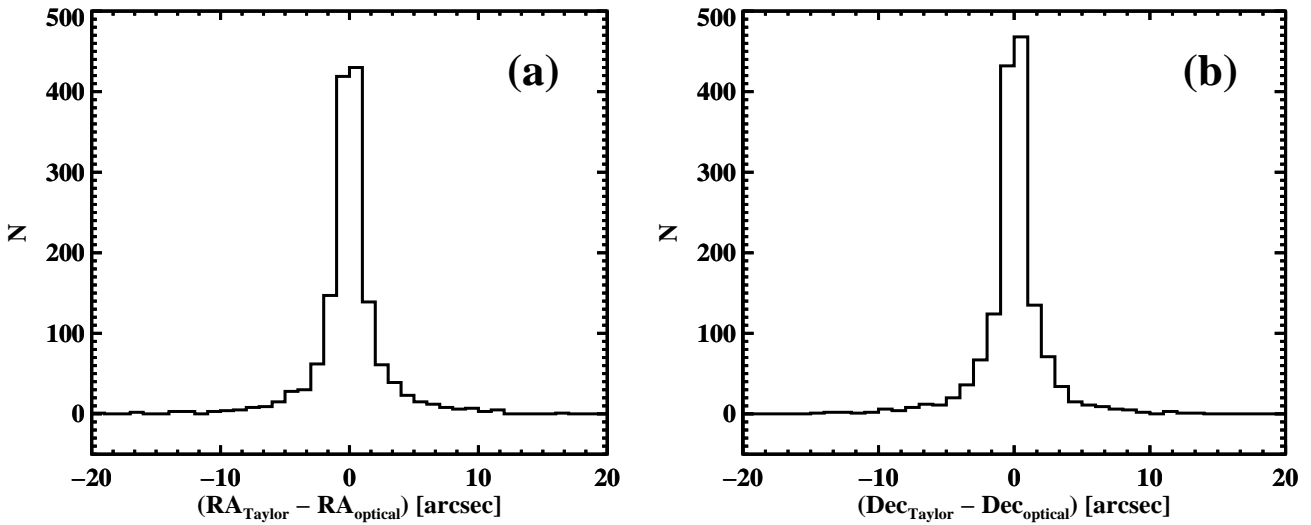,width=\textwidth}}
\caption{Histograms showing the distribution of radio and optical positional
offsets in our RM-redshift catalog in (a) Right Ascension and (b)
Declination. The catalog includes a small number of associations with
offsets outside the range shown (these sources, like all those beyond an
offset of $\approx7''$, have a double-lobed or complex radio morphology,
leading to large separations between radio and optical source
positions).\label{fig:offsets}}
\end{figure}

\clearpage

\begin{figure}
\centerline{\psfig{file=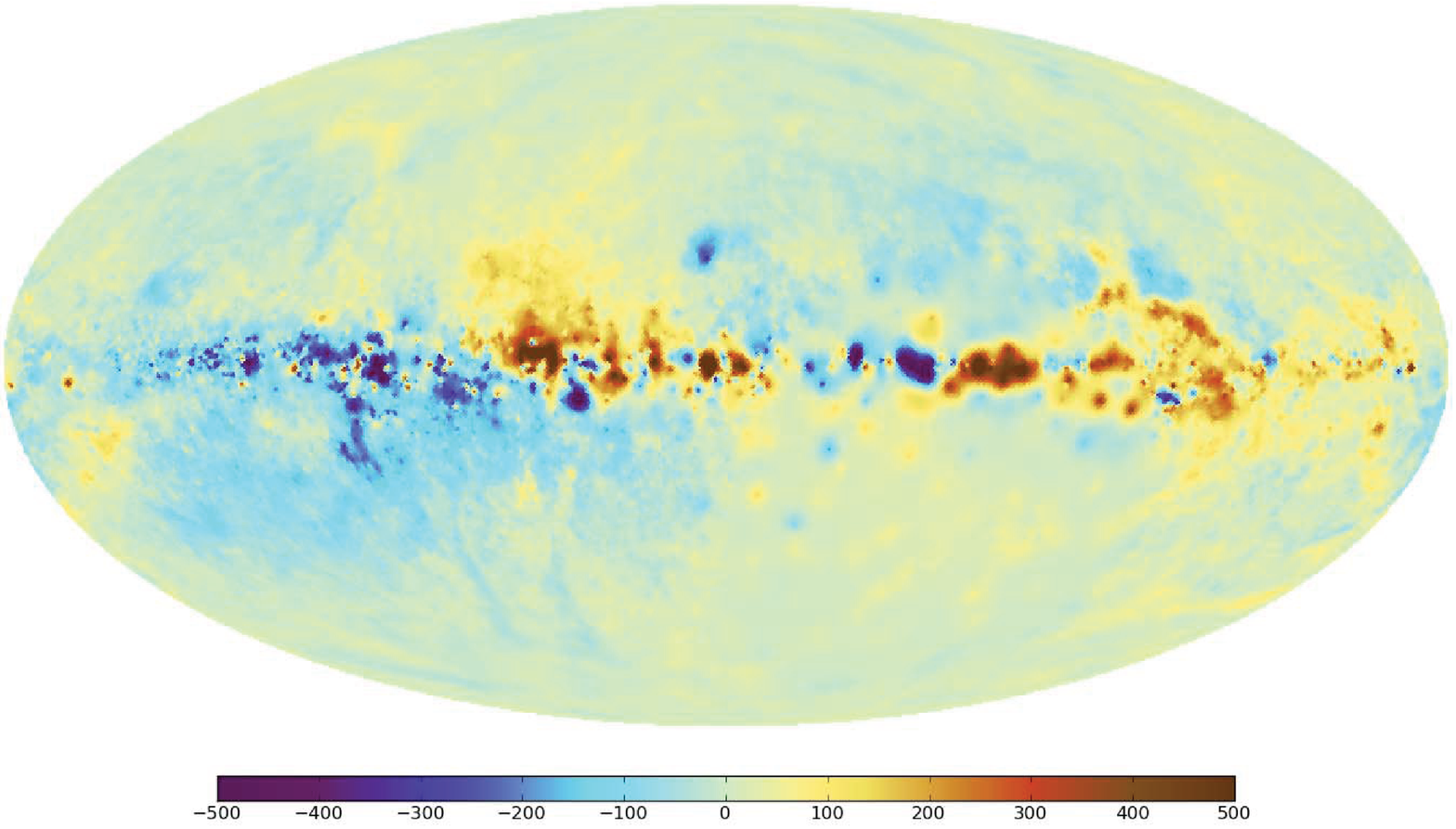,width=\textwidth}} 
\caption{Map of the foreground Faraday sky in rad m$^{-2}$ by
\citet{oppermann}.\label{fig:foreground_niels}}
\end{figure}

\begin{figure}
\centerline{\psfig{file=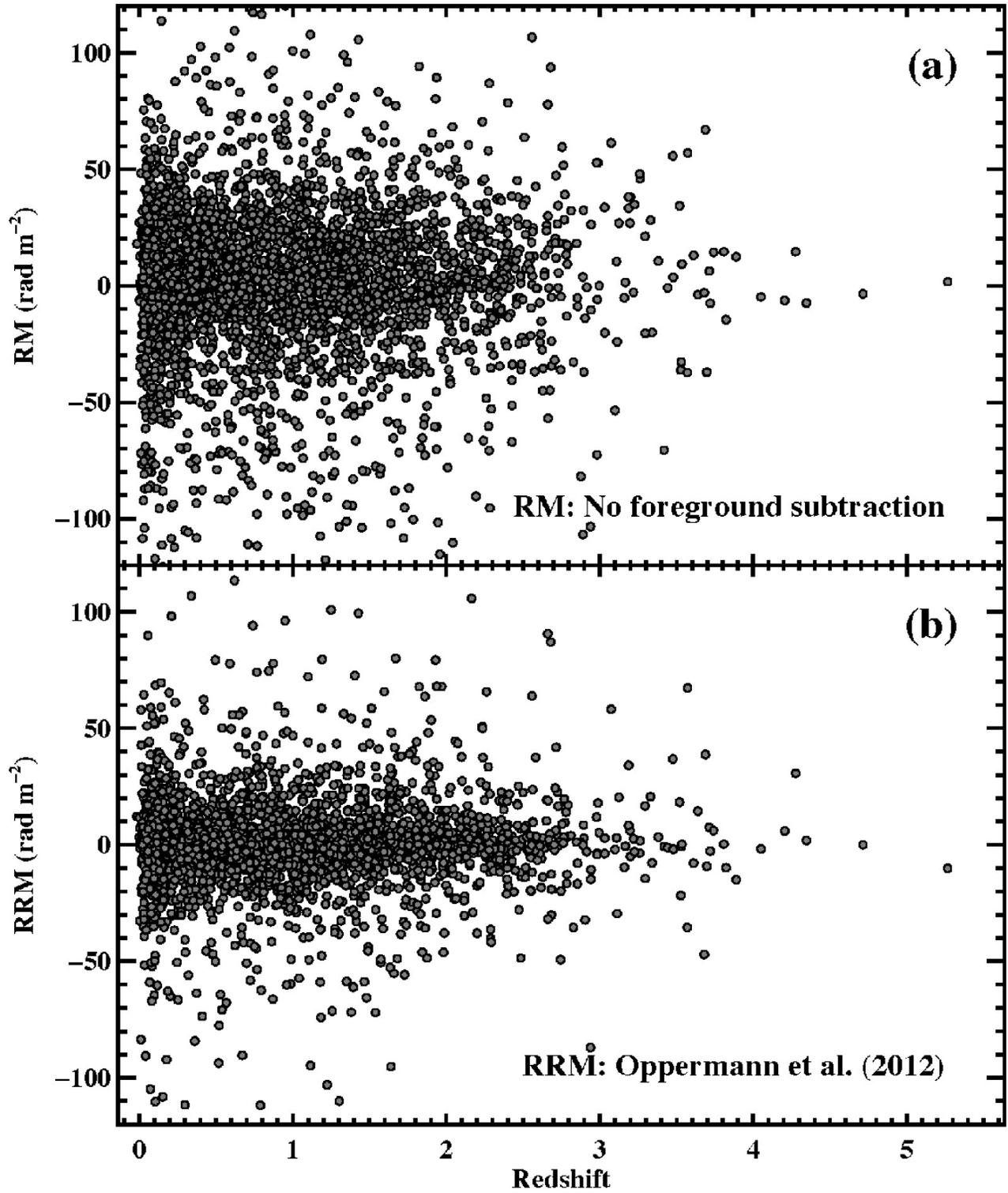,width=\textwidth}} 
\caption{Comparison of (a) RM vs.\ $z$ and (b) RRM vs.\ $z$, showing the impact
of the GRM subtraction.\label{fig:redshift_methods}}
\end{figure}

\begin{figure}
\centerline{\psfig{file=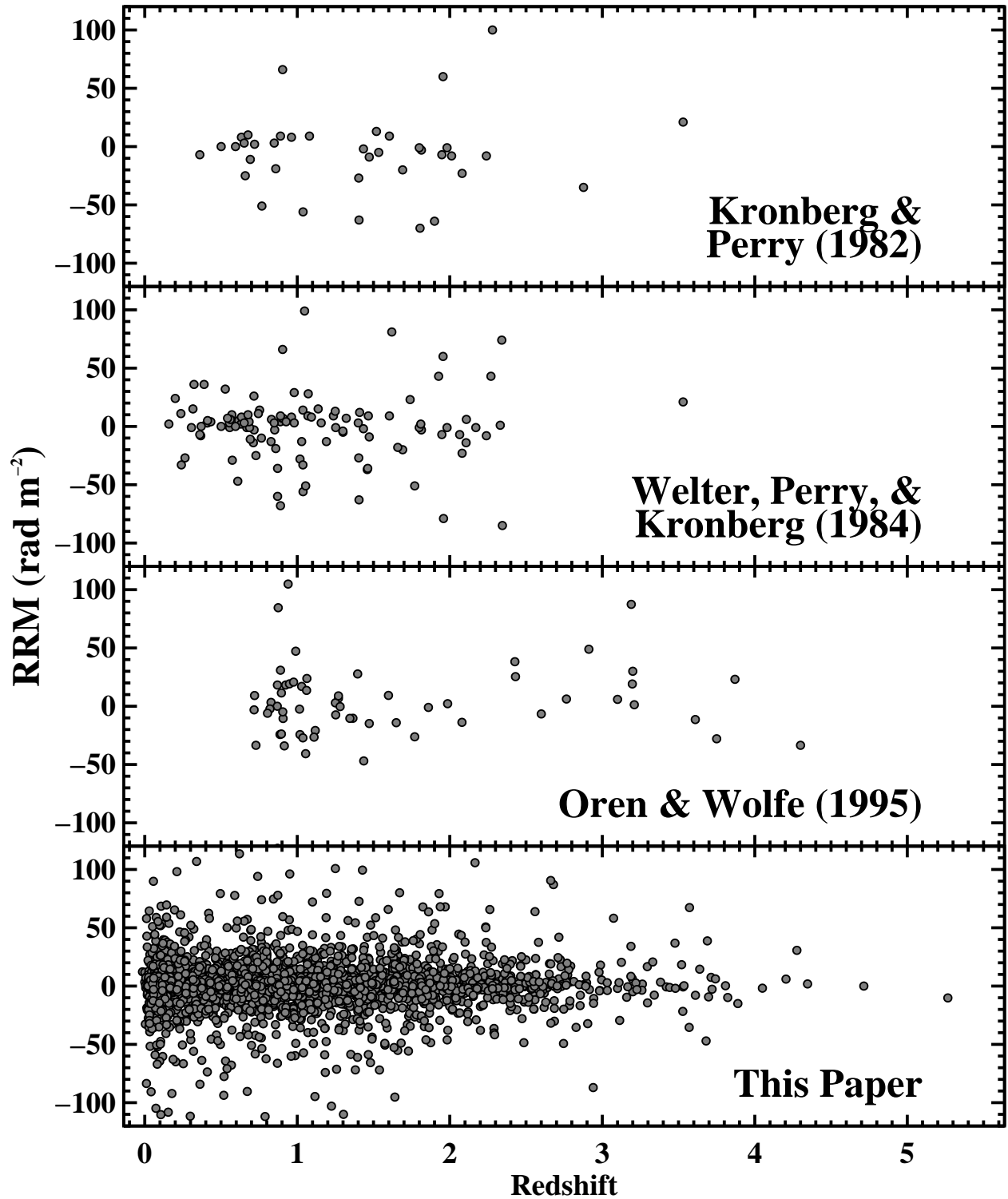,width=\textwidth}} 
\caption{A comparison of our RRM-redshift catalog (bottom panel) with
previously published data sets. A catalog of 268 objects discussed by
\citet{kronberg2008} has not yet been made available.
\label{fig:catalog_comparison}}
\end{figure}

\begin{figure}
\centerline{\psfig{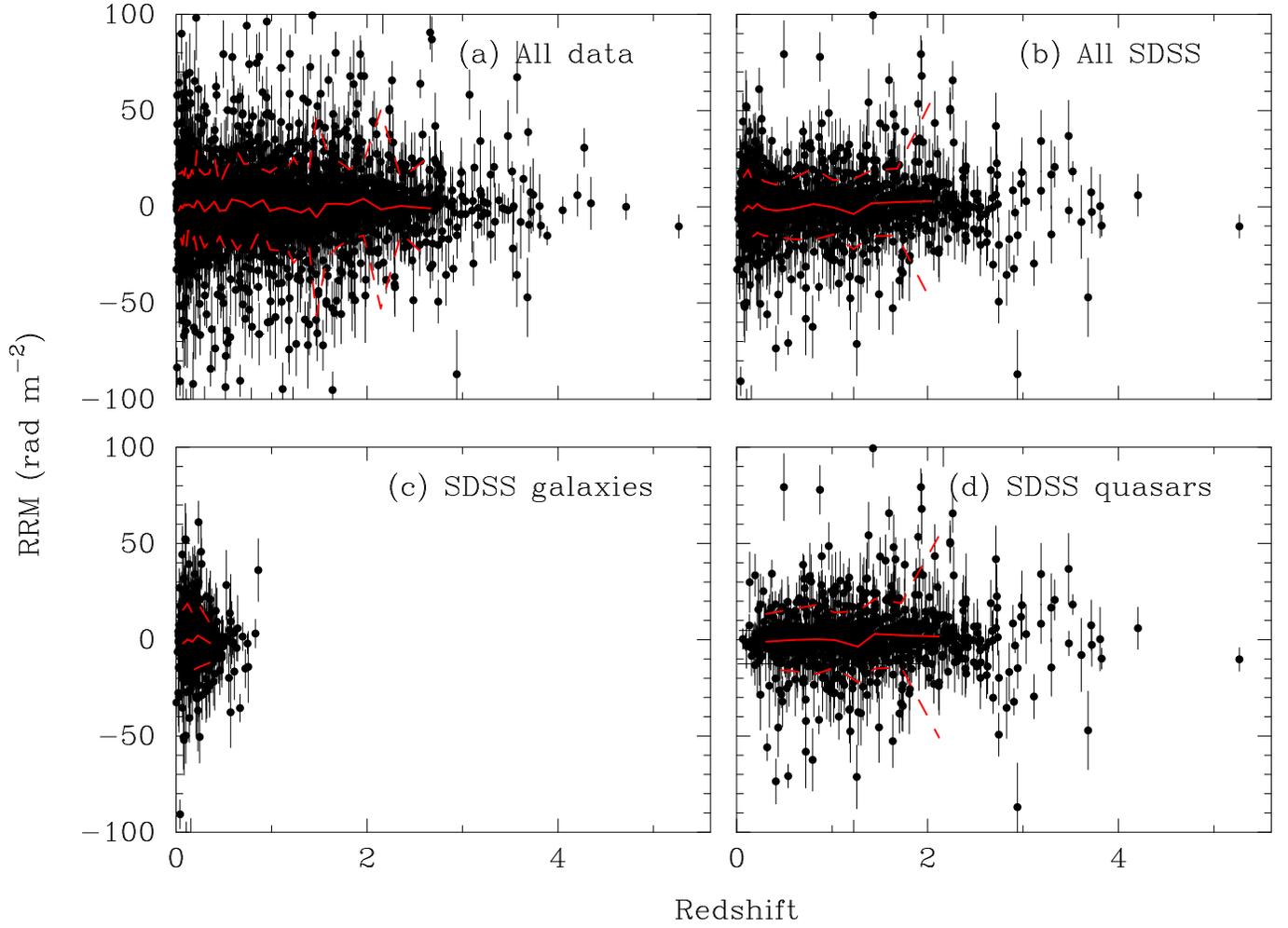}}
\caption{RRM vs.\ redshift for (a) the whole RRM-redshift catalog of 3650
sources, (b) the subset of 1376 sources with selected redshifts from the
SDSS, (c) the 516 galaxies with selected redshifts from the SDSS, and (d)
the 860 quasars with selected redshifts from the SDSS. In each panel,
individual data points are shown in black, while the mean RRM and
$\pm1\sigma$ values of RRM on either side of this mean in independent
adjacent bins of 50 points are shown by the solid red line and the dashed
red lines, respectively. Note that there are 15 sources with
RRM $<-100$~rad~m$^{-2}$ and 10 sources with RRM $>+100$~rad~m$^{-2}$
that are outside the range plotted.}
\label{fig_rrm_z}
\end{figure}

\begin{figure}
\centerline{\psfig{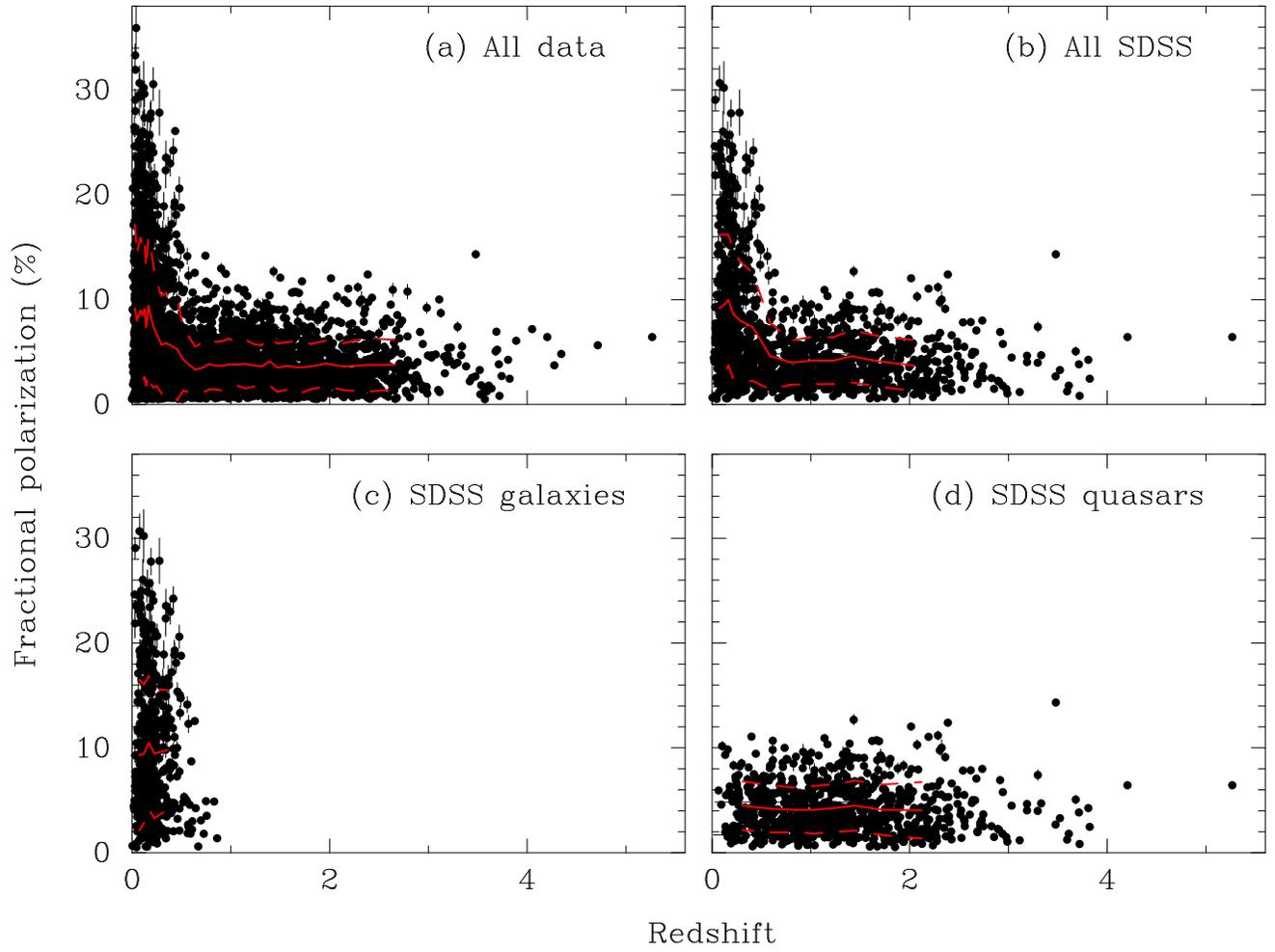}}
\caption{As for Figure~\ref{fig_rrm_z}, but showing fractional
polarization vs.\ redshift.}
\label{fig_p_z}
\end{figure}

\begin{figure}
\centerline{\psfig{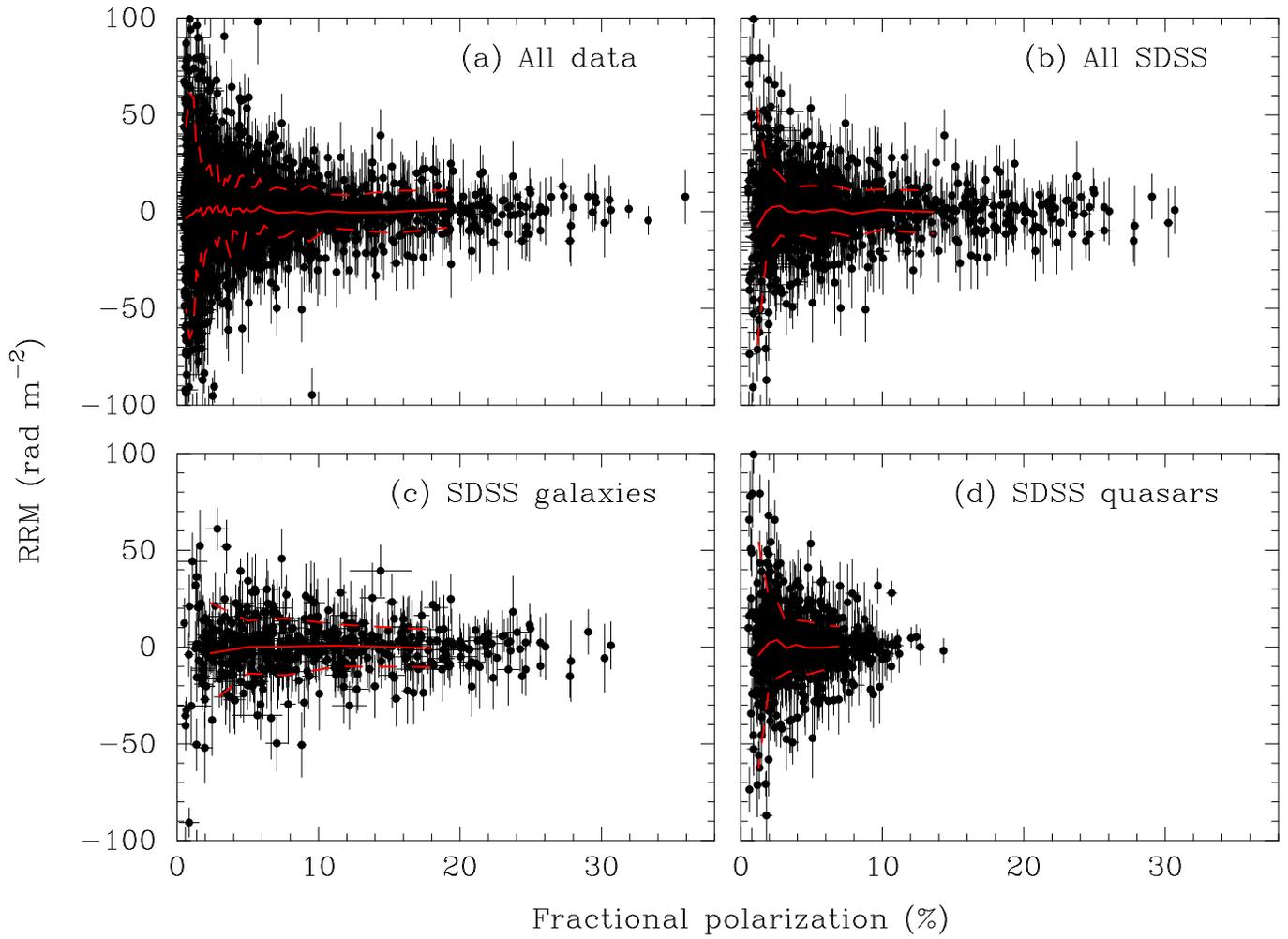}}
\caption{As for Figure~\ref{fig_rrm_z}, but showing RRM vs.\ fractional
polarization. There are 15 sources with
RRM $<-100$~rad~m$^{-2}$ and 10 sources with RRM $>+100$~rad~m$^{-2}$
that are outside the range plotted.}
\label{fig_rrm_p}
\end{figure}

\end{document}

%% file: rm_z_catalog.tex
\begin{deluxetable}{lccccccccccccccccc}
\small
\tablecolumns{14}
\tablewidth{0pt}
\tablecaption{Selected Columns from Our RM-Redshift Catalog of Polarized
Radio Sources with Optical Counterparts \label{tab:rm_z_catalog}}
\tabletypesize{\footnotesize}
\tablehead{
\colhead{NVSS ID}      & \multicolumn{2}{c}{SDSS}        &&
  \multicolumn{2}{c}{2dFGRS}      && \multicolumn{2}{c}{6dFGS}
  && \multicolumn{2}{c}{2QZ/6QZ}     & \multicolumn{1}{c}{NED} &
  \multicolumn{1}{c}{SIMBAD} && \multicolumn{2}{c}{Selected Value} & \multicolumn{1}{c}{RM}
\\
\cline{2-3} \cline{5-6} \cline{8-9} \cline{11-12} \cline{16-17}
\\[-2ex]
 & \colhead{$z$} & \colhead{Class} && \colhead{$z$} &
 \colhead{Class} && \colhead{$z$} & \colhead{Class\tablenotemark{a}} &&
 \colhead{$z$} & \colhead{Class} &           \colhead{$z$} &
 \colhead{$z$} && \colhead{Source} &  \colhead{$z$} & \colhead{(${\rm rad~m^{-2}}$)}
\\
     \colhead{(1)} & \colhead{(2)} &   \colhead{(3)} && \colhead{(4)} &
     \colhead{(5)} && \colhead{(6)} &                    \colhead{(7)} &&
     \colhead{(8)} &   \colhead{(9)} &          \colhead{(10)} &
     \colhead{(11)} &&   \colhead{(12)} &  \colhead{(13)} & \colhead{(14)}
}
\startdata
J000010$+$305559 & \nodata & \nodata && \nodata & \nodata && \nodata &
\nodata && \nodata & \nodata & 1.801 & 1.8 &&    NED    &    1.801
& $   -37.9 \pm 11.0$
\\
J000030$-$112119 & \nodata & \nodata && \nodata & \nodata && 0.10285 &
A(iv) && \nodata & \nodata & \nodata & \nodata &&    6dFGS  &    0.10285
 & $    +6.4 \pm 12.4$
\\
J000132$+$145609 & 0.39878 &  B && \nodata & \nodata && \nodata & \nodata
&& \nodata & \nodata & 0.39892 & \nodata &&    SDSS   &    0.39878
 & $   -34.9 \pm  3.8$ 
\\
J000153$-$302508 & \nodata & \nodata && \nodata & \nodata && 1.30252 &    B
&& \nodata & \nodata & 1.30252 & \nodata &&    6dFGS  &    1.30252
 & $    -0.1 \pm  6.2$
\\
J000154$+$020453 & \nodata & \nodata && \nodata & \nodata && \nodata &
\nodata && \nodata & \nodata & 0.402 & \nodata &&    NED    &    0.402
& $   -14.9 \pm 10.5$
\\
J000255$-$265451 & \nodata & \nodata && 0.0667 &    B    && 0.06666 &    B
&& \nodata & \nodata & 0.06659 & 0.0665 &&    6dFGS  &    0.06666
 & $    +0.8 \pm  7.2$
\\
J000322$-$172711 & \nodata & \nodata && \nodata & \nodata && \nodata &
\nodata && \nodata & \nodata & 1.465 & \nodata &&    NED    &    1.465
 & $   -33.2 \pm  2.4$ 
\\
J000325$-$272631 & \nodata & \nodata && 0.2501 &    B    && \nodata &
\nodata && \nodata & \nodata & \nodata & \nodata &&    2dFGRS &    0.2501
& $    +1.9 \pm 12.3$ 
\\
J000327$-$154706 & \nodata & \nodata && \nodata & \nodata && \nodata &
\nodata && \nodata & \nodata & 0.508 & 0.508 &&    NED    &    0.508
 & $    -2.8 \pm  4.0$
\\
J000342$-$115149 & \nodata & \nodata && \nodata & \nodata && 1.30999 &    B
&& \nodata & \nodata & 1.30999 & \nodata &&    6dFGS  &    1.30999
 & $    -3.8 \pm 13.3$

\enddata

\tablenotetext{a}{The high proportion of 6dFGS and B class detections
  compared to the catalog as a whole is due to the low RAs of this sample
  (cf. Figure \ref{fig:sky coverage}.)\vspace{-4ex}}

\tablecomments{Table~\ref{tab:rm_z_catalog} is published in its entirety in the
  electronic edition of {\em The Astrophysical Journal Supplement Series}.  The first 10 rows are
shown here for
  guidance regarding its form and content.}

\normalsize
\end{deluxetable}

%% file: table4.tex
\begin{deluxetable}{lrrrrccrrr}
\tablecolumns{10}
\tablewidth{0pt}
\tablecaption{The contents of our RM-redshift catalog, compiled from four optical
surveys and two
databases, broken down by association class. \label{tab:sdss} }
\tabletypesize{\scriptsize}
\tablehead{\colhead{Class}\hspace{6ex} & \colhead{SDSS} & \colhead{2dFGRS} &
\colhead{6dFGS} & \colhead{2QZ/6QZ} & \colhead{NED} & \colhead{SIMBAD} &
\colhead{All} & \colhead{Selected} & \colhead{Selected,} \\
 & & & & & & & & & \colhead{$|b| \ge 20^\circ$}}
\startdata
A\dotfill &      1135 &        15 &       23 &       27 & \nodata & \nodata
&      1200  & 1162 & 1160 \\
~~A(i)\dotfill &      528 &        6 &       11 &       16 & \nodata &
\nodata &      561 & 539 & 538 \\
~~A(ii)\dotfill &       12 &        0 &        0 &        1 & \nodata &
\nodata &       13 & 12 & 12 \\
~~A(iii)\dotfill &      225 &        3 &        5 &        3 & \nodata &
\nodata &      236 & 231 & 231 \\
~~A(iv)\dotfill &       83 &        2 &        2 &        2 & \nodata &
\nodata &       89 & 86 & 86 \\
~~A(v)\dotfill &      220 &        4 &        3 &        5 & \nodata &
\nodata &      232  & 225 & 224 \\
~~A(vi)\dotfill &       17 &        0 &        2 &        0 & \nodata &
\nodata &       19  & 19 & 19 \\
~~A(vii)\dotfill &       50 &        0 &        0 &        0 & \nodata &
\nodata &       50 & 50 & 50 \\
B\dotfill &       77 &       43 &      236 &       35 & \nodata & \nodata &
391 & 375 & 345 \\
C\dotfill &      174 &        9 &       19 &        3 & \nodata & \nodata &
205  & 195 & 193 \\
D\dotfill & \nodata & \nodata & \nodata & \nodata &     3186 & \nodata &
3186 & 2023 & 1742 \\
E\dotfill & \nodata & \nodata & \nodata & \nodata & \nodata &     1907 &
1907 & 248 & 210 \\
Total\dotfill &     1386 &       67 &      278 &       65 &     3186 &
1907 &     6889 & 4003 & 3650 \\
\tableline
Selected\dotfill &     1380 &       44 &      264 &       44 &     2023 &
248 &     4003 & N/A & N/A \\
Selected, $|b|\ge20^\circ$\dotfill &     1376 &       44 &      234 &
44 &     1742 &      210 &     3650 & N/A & N/A 
\enddata
\tablecomments{The row labelled `Total' gives the total number of
associations made between \citet{taylor} sources and the optical survey or
database in question, while the rows labelled `Selected' indicate the number
of cases where this survey or database is the source of the ``selected
redshift'' (see \S \ref{subsec:bestz})  --- for example, there are many
cases for which the same \citet{taylor} source has a redshift in both SDSS
and NED, in which case the SDSS redshift is selected over the NED one.  The
column labelled `All' lists the total number of times a given association
class is assigned to a match between a  \citet{taylor} source and an optical
counterpart, while the columns  labelled `Selected' indicate the number of
selected redshifts in each  association class --- for example, most SIMBAD
matches (class~E) also have another match in class A, B or C from an optical
survey.}
\end{deluxetable}

%% file: rrm_z_catalog.tex
\begin{deluxetable}{lrrccccccccc}
\setlength{\tabcolsep}{0.5ex}
\tablecolumns{13}
\tablewidth{0pt}
\tablecaption{Selected Columns from Our RRM-Redshift Catalog of Polarized
Radio Sources with Optical Counterparts  \label{tab:rrm_z_catalog}}
\tabletypesize{\scriptsize}
\tablehead{
\colhead{NVSS ID}                    &      \colhead{$\ell$} &
\colhead{$b$} & \colhead{NVSS RM} & \colhead{Stokes $I$ Flux} & 
\colhead{Frac Pol} & \colhead{Source} &  \colhead{Class} &
\colhead{Object} & \colhead{Redshift}                   &                 \colhead{GRM} &                 \colhead{RRM} \\
%\cline{9-10}\\[-4.5ex]
  & \colhead{($^{\circ}$)} & \colhead{($^{\circ}$)} & \colhead{(${\rm
rad~m^{-2}}$)} &      \colhead{(mJy)} &  
\colhead{(\%)} &  & & \colhead{Type} & &  \colhead{(${\rm rad~m^{-2}}$)} & \colhead{(${\rm rad~m^{-2}}$)} \\
     \colhead{(1)} &         \colhead{(2)} &         \colhead{(3)} &
\colhead{(4)} &        \colhead{(5)} &      \colhead{(6)} &
\colhead{(7)} &      \colhead{(8)} &    \colhead{(9)} &   \colhead{(10)} &
\colhead{(11)} &      \colhead{(12)} 
}

\startdata
              J000010$+$305559 & $   110.15$ & $   -30.66$ & $   -37.9 \pm
11.0$ & $    88.2 \pm     2.7$     & $     6.5 \pm
0.3$ & NED & D & Quasar & $1.801 \pm  0.007$ &    $ -74.2 \pm     7.5$ & $ +36.3 \pm    13.3$
\\
              J000030$-$112119 & $    83.29$ & $   -70.20$ & $    +6.4 \pm
12.4$ & $    80.9 \pm     2.9$ &  $     6.8 \pm
0.4$ & 6dFGS & A(iv) & ... & $0.10285 \pm ... $ &    $  +1.9 \pm     4.1$ & $  +4.5 \pm    13.1$
\\
              J000132$+$145609 & $   105.37$ & $   -46.23$ & $   -34.9 \pm
3.8$ & $   314.6 \pm    11.1$ & $     5.5 \pm
0.1$ & SDSS & B & Quasar & $0.39878 \pm ... $ &     $ -28.9 \pm     5.2$ & $  -6.0 \pm     6.4$
\\
              J000153$-$302508 & $    13.14$ & $   -78.66$ & $    -0.1 \pm
6.2$ & $   173.8 \pm     5.2$ &  $     6.4 \pm
0.2$ & 6dFGS & B & ... & $1.30252 \pm ...$ &    $ +15.4 \pm     4.1$ & $ -15.5 \pm     7.4$
\\
              J000154$+$020453 & $    98.76$ & $   -58.45$ & $   -14.9 \pm
10.5$ & $   300.7 \pm     9.7$ & $     3.3 \pm
0.2$ & NED & D & Galaxy & $0.402 \pm ...$ &    $  -8.3 \pm     5.4$ & $  -6.6 \pm    11.8$
\\
              J000255$-$265451 & $    31.30$ & $   -79.20$ & $    +0.8 \pm
7.2$ & $    94.4 \pm     3.2$ &  $    12.8 \pm
0.4$ & 6dFGS & B & ... & $0.06666 \pm ...$ &     $  +4.9 \pm     3.9$ & $  -4.1 \pm     8.2$
\\
              J000322$-$172711 & $    71.53$ & $   -75.28$ & $   -33.2 \pm
2.4$ & $  2414.8 \pm    72.4$ & $     1.5 \pm
0.0$ & NED & D & Quasar & $1.465 \pm  0.003$ &      $  +1.9 \pm     4.6$ & $ -35.1 \pm     5.2$
\\
              J000325$-$272631 & $    28.49$ & $   -79.33$ & $    +1.9 \pm
12.3$ & $   144.0 \pm     5.1$ & $     5.2 \pm
0.3$ & 2dFGRS & B & ... & $0.2501 \pm ...$ &    $  +5.9 \pm     4.1$ & $  -4.0 \pm    13.0$
\\
              J000327$-$154706 & $    76.02$ & $   -74.10$ & $    -2.8 \pm
4.0$ & $   527.3 \pm    15.8$ &  $     3.2 \pm
0.1$ & NED & D & Galaxy & $0.508 \pm ...$ &    $  -5.4 \pm     3.4$ & $  +2.6 \pm     5.3$
\\
              J000342$-$115149 & $    84.35$ & $   -71.07$ & $    -3.8 \pm
13.3$ & $   351.7 \pm    12.2$ & $     1.6 \pm
0.1$ & 6dFGS & B & ... & $1.30999 \pm  0.0068$ &  $  -2.4 \pm     3.3$ & $  -1.4 \pm    13.7$
\enddata
\tablecomments{Table~\ref{tab:rrm_z_catalog} is published in its entirety in
the
  electronic edition of {\em The Astrophysical Journal Supplement Series}.
The first 10 rows are
shown here for
guidance regarding its form and content.}
\end{deluxetable}